# In silico comparison of SARS-CoV-2 spike protein-ACE2 binding affinities across species and implications for viral origin

Sakshi Piplani[1,2], Puneet Kumar Singh[2], David A. Winkler[3-6]*, Nikolai Petrovsky[1,2]*

[1] College of Medicine and Public Health, Flinders University, Bedford Park 5046, Australia

[2] Vaxine Pty Ltd, 11 Walkley Avenue, Warradale 5046, Australia

[3] La Trobe University, Kingsbury Drive, Bundoora 3086, Australia

[4] Monash Institute of Pharmaceutical Sciences, Monash University, Parkville 3052, Australia

[5] School of Pharmacy, University of Nottingham, Nottingham NG7 2RD. UK

[6] CSIRO Data61, Pullenvale 4069, Australia

*Joint senior authors

Corresponding authors: –

Prof. Nikolai Petrovsky: nikolai.petrovsky@flinders.edu.au

Prof. Dave Winkler: d.winkler@latrobe.edu.au

# Abstract

The devastating impact of the COVID-19 pandemic caused by SARS–coronavirus 2 (SARS-CoV-2) has raised important questions about viral origin, mechanisms of zoonotic transfer to humans, whether companion or commercial animals can act as reservoirs for infection, and why there are large variations in SARS-CoV-2 susceptibilities across animal species. Powerful *in silico* modelling methods can rapidly generate information on newly emerged pathogens to aid countermeasure development and predict future behaviours.

Here we report an *in silico* structural homology modelling, protein-protein docking, and molecular dynamics simulation study of the key infection initiating interaction between the spike protein of SARS-Cov-2 and its target, angiotensin converting enzyme 2 (ACE2) from multiple species. Human ACE2 has the strongest binding interaction, significantly greater than for any species proposed as source of the virus. Binding to pangolin ACE2 was the second strongest, possibly due to the SARS-CoV-2 spike receptor binding domain (RBD) being identical to pangolin CoV spike RDB. Except for snake, pangolin and bat for which permissiveness has not been tested, all those species in the upper half of the affinity range (human, monkey, hamster, dog, ferret) have been shown to be at least moderately permissive to SARS-CoV-2 infection, supporting a correlation between binding affinity and permissiveness. Our data indicates that the earliest isolates of SARS-CoV-2 were surprisingly well adapted to human ACE2, potentially explaining its rapid transmission.

# Introduction

The devastating impact of COVID-19 infections caused by SARS–coronavirus 2 (SARS-CoV-2) has stimulated unprecedented international activity to discover effective vaccines and drugs for this and other pathogenic coronaviruses.[1-4] It has also raised important questions about the mechanisms of zoonotic transfer of viruses from animals to humans, whether companion animals or those used for commercial purposes can act as reservoirs for infection, and the reasons for the large variations in SARS-CoV-2 susceptibility across animal species.[5-7] Understanding how viruses move between species may help us prevent or minimize similar events in the future. Methods that elucidate the molecular basis for differences in species susceptibilities of may also shed light on why different human sub-groups exhibit differences in susceptibilities.[8]

The most important features of SARS-CoV-2 are its spike protein (S protein) and a functional polybasic cleavage site at the S1–S2 boundary.[9] The SARS-CoV-2 spike monomer consists of a fusion peptide, two heptad repeats, an intracellular domain, N-terminal domain, two subdomains and a transmembrane region.[10] The angiotensin converting enzyme 2 (ACE2) was identified as the main receptor for the SARS-CoV-2 S protein, as it is for SARS CoV, binding to which is a critical initiating event for infection. ACE2 is relatively ubiquitous in humans, existing in cell membranes in the lungs, arteries, heart, kidney, and intestines. It consists of an N-terminal peptidase M2 domain and a C-terminal collectrin renal amino acid transporter domain.

Non-human species vary markedly in their susceptibility to SARS-Cov-2[7,11,12] and their ACE2 receptor binding domains also differ substantially. The phylogenetic tree showing relatedness of ACE2 proteins across selected animal species is illustrated in Supplementary Figure 1. Several species, notably pangolin, bat and snake, have been postulated as a source or intermediate host for SARS-Cov-2.[11,13,14]

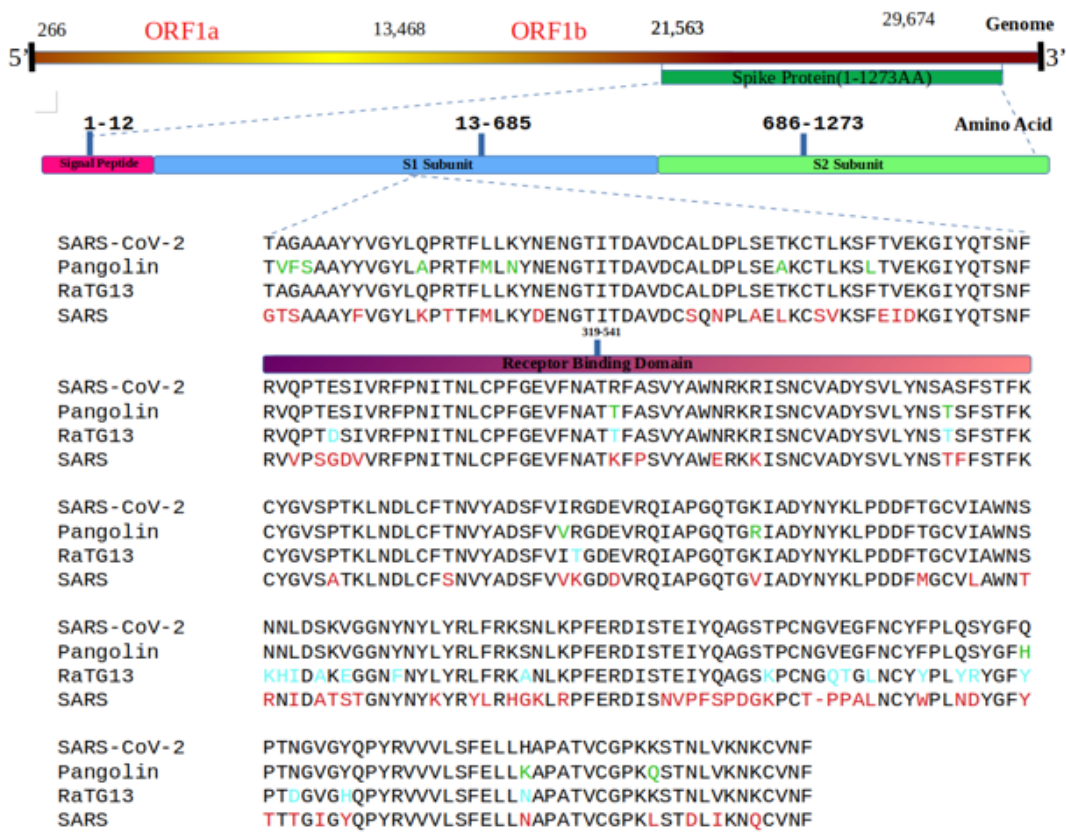

**Figure 1**. Sequence alignment of four closely related spike proteins. Difference in pangolin and bat spike proteins colored cyan and differences in SAR spike denoted in red. The receptor binding domain (RBD) is denoted by the red bar in the sequence alignments

We and others[15,16] postulate that variation in ACE2 structures between species will determine the binding strength of SARS-CoV-2 S protein and suggest which species are permissive to SARS-CoV-2 infection. For example, the low binding affinity of SARS S protein for mouse ACE2 has been postulated as the reason mice are largely non-permissive to SARS infection. Direct measurement of the binding affinity of SARS-CoV-2 S protein to ACE2 (e.g. using cell lines transfected with ACE2 proteins from different species) would be very useful but is time consuming, and purified or recombinant ACE2 proteins from all relevant animal species are not yet available.

Here we show how fast, efficient *in silico* structural modelling and docking algorithms from structure-based drug design can be used to determine the relative binding affinities of the SARS-CoV-2 S protein for ACE2, across multiple common and exotic animal species.[17-19] We provide novel insights into the species-specific nature of this interaction and impute which species might be permissive for SARS-CoV-2 that may help elucidate the origin of SARS-CoV-2 and the mechanisms for its zoonotic transmission.

**Results and Discussion**

The results of docking the receptor binding domains (RBDs) of SARS-CoV-2 S and ACE2 receptors of various species using the HDOCK server, refined by MD simulations, are summarized in Tables 1 and 2. The calculated binding energies for the interactions of SARS-CoV-2 with ACE2 from the species studied are summarized in Table 2, and the MMPBSA binding energies are listed for comparison. The energies calculated by equation 1 and those from the MMPBSA algorithm (see Materials and Methods) correlated strongly ($r^2=0.76$) but the energies correlated poorly with sequence similarity ($r^2=0.27$). The Kendall tau rank correlation for the two different methods of calculating binding energies was 0.94. Also shown is the observed and predicted SARS-CoV-2 susceptibilities of the species from analysis of phylogenetic clustering and sequence alignment with currently known ACE2s utilized by SARS-CoV-2, as described by Qiu et al.[20] (Supplementary Figure 4) For comparison, we show the interaction energies of the S and ACE2 proteins for each species calculated by Wu et al. using an automatic docking method, ICM-Pro[21] (no subsequent MD refinement), and observational *in vivo* data on SARS-CoV-2 infectivity and disease symptoms in the species, where this has been reported. The interaction energies calculated by Wu et al., based only on docking calculations, predict that monkey, hamster, tiger and bat have higher affinity than human ACE2 for the spike protein which is clearly not consistent with the known infectivities of the species in Table 2. This suggests that more sophisticated computational methods are required to obtain binding affinities more consistent with observation, as in the work presented here and by others, e.g.[22]

While this paper was being prepared, Rodrigues et al. published a similar study that used the docking method HADDOCK to estimate the relative strength of binding affinities of SARS-Cov-2 spike protein for ACE2 proteins of 30 species.[22] The docking experiments were followed by short MD simulations. This study found 18 species had higher affinity than the human ACE2, including dog, pangolin, ferret, Siberian tiger, horseshoe bat, civet, hamster and guinea pig but also goldfish, sheep, cat, horse and rabbit. Goldfish (no known permissiveness of infection with SARS-Cov-2) had the second highest affinity after dog. The authors stated that, despite their method having some discrimination between species that are susceptible and those that are not, their predictions were not entirely correct. For example, they rank guinea pig ACE2 (SARS-CoV-2 negative) as a better receptor for SARS-CoV-2 RBD than for human, cat, horse, or rabbit ACE2 (all SARS-CoV-2 positive species), despite experiments showing that there is negligible binding between the two proteins.[23] One explanation may be that their MD simulations were short (simulation time not listed) and it longer simulations like those used in this study are needed to obtain more accurate ranking of the binding affinities.

Other relevant papers were also published while this paper was in preparation and review. Recently, Damas et al. published an analysis of ACE2 sequences from 410 vertebrate species, including 252 mammals, to study the probability that ACE2 could be used as a receptor by SARS-CoV-2.[15] They classed the species into five risk groups. Man, apes, and monkeys were in the very high-risk group while Chinese hamsters, whales and porpoises were prominent examples of species in the high-risk group.

Note added in proof: While this paper was undergoing review, a highly relevant paper was published in Scientific Reports (5 Oct) by Lam et al.[24] on the species specificity of spike-ACE2 interaction. This provides validation of our computational approach and results we obtained on the relative binding affinities of non-human species. As with our study they used MODELLER to generate ACE2 structures for difference species and selected the best refined model using DOPE scores. They calculated free energy differences, $\Delta\Delta G$, between human ACE2 binding and that of non-human species ACE2 to SARS-Cov-2 spike using free energy perturbation methods. However, energy differences they calculated are similar to those we calculated. Unfortunately, Lam et al. did not report the $\Delta\Delta G$ for the most important species, pangolin, making our study significant in reporting this binding energy.

**Table 1**. ACE2 RBD residues interacting with the S protein RBD from MD simulations of complexes. Residues interacting with the same S residue in different species ACE2, that differ from those in human ACE2, are labelled green (conservative replacements) or red (non-conservative replacements).

| Species | Accession Number | Position | | | | | | | | | | | | | | | | % common residues |
|---|---|---|---|---|---|---|---|---|---|---|---|---|---|---|---|---|---|---|
| | | 19 | 24 | 27 | 28 | 30 | 31 | 34 | 37 | 38 | 41 | 42 | 79 | 83 | 330 | 353 | 393 | |
| *Homo sapiens* (human) | Q9BYF1 | S | Q | T | F | D | K | H | E | D | Y | Q | L | Y | N | K | R | 100 |
| *Macaca fascicularis* (monkey) | A0A2K5X283 | S | Q | T | F | D | K | H | E | D | Y | Q | L | Y | N | K | R | 100 |
| *Panthera tigris* (tiger) | XP_007090142.1 | S | L | T | F | D | K | H | E | E | Y | Q | L | Y | K | K | R | 94 |
| *Bos Taurus* (cow) | NP_001019673.2 | S | Q | T | F | E | K | H | E | D | Y | Q | M | Y | N | K | R | 88 |
| *Mesocricetus auratus* (hamster) | A0A1U7QTA1 | S | Q | T | F | D | L | Q | E | D | Y | Q | L | Y | N | K | R | 88 |
| *Felis catus* (cat) | Q56H28 | S | L | T | F | E | K | H | E | E | Y | Q | L | Y | N | K | R | 81 |
| *Rhinolophus sinicus* (bat) | U5WHY8 | S | E | M | F | D | K | T | E | D | H | Q | L | Y | N | K | R | 75 |

| Species | Accession Number | Position | | | | | | | | | | | | | | | | % common residues |
|---|---|---|---|---|---|---|---|---|---|---|---|---|---|---|---|---|---|---|
| | | 19 | 24 | 27 | 28 | 30 | 31 | 34 | 37 | 38 | 41 | 42 | 79 | 83 | 330 | 353 | 393 | |
| *Homo sapiens (human)* | Q9BYF1 | S | Q | T | F | D | K | H | E | D | Y | Q | L | Y | N | K | R | 100 |
| *Paguma larvata (civet)* | Q56NL1 | S | L | T | F | E | K | Y | E | Q | Y | Q | L | Y | N | K | R | 75 |
| *Equus ferus caballus (horse)* | F6V9L3 | S | L | T | F | D | K | S | E | E | H | Q | L | Y | N | K | R | 75 |
| *Mustela putorius furo (ferret)* | Q2WG88 | D | L | T | F | E | K | T | E | E | Y | Q | - | Y | N | K | R | 69 |
| *Canis luparis (dog)* | J9P7Y2 | - | L | T | F | E | K | Y | E | E | Y | Q | L | Y | N | K | R | 69 |
| *Mus musculus (mouse)* | Q8R0I0 | S | N | T | F | N | N | Q | E | D | Y | Q | Y | F | N | K | R | 63 |
| *Manis javanica (pangolin)* | XP_017505752.1 | - | E | T | F | E | K | S | E | E | Y | Q | I | Y | N | K | R | 63 |
| *Ophiophagus Hannah (snake)* | ETE61880.1 | Q | V | K | F | E | Q | A | - | D | Y | N | N | F | N | L | R | 38 |

**Table 2.** Binding energies of SARS-Cov-2 spike to ACE2 for different species and potential species susceptibilities from other studies

| Species | $E_{eqn1}$ (kcal/mol) | $E_{MMPBSA}$ (kcal/mol) | SARS-Cov-2 infectivity | Binding score Wu et al.[21] | ACE2 used by SAR-Cov-2 spike[20] |
|---|---|---|---|---|---|
| *Homo sapiens* (human) | -52.8 | -57.6 ± 0.25 | Permissive, high infectivity, severe disease in 5-10%, | -50.1 | confirmed |
| *Manis javanica* (pangolin) | -52.0 | -56.3 ± 0.4 | Permissive [25,26] | -46.4 | predicted |
| *Canis luparis* (dog) | -50.8 | -49.5 | Permissive, low/mod infectivity, no overt disease [27,28] | -40.7 | predicted |
| *Macaca fascicularis* (monkey) | -50.4 | -50.8 | Permissive, high infectivity, lung disease [11] | -51.5 | |
| *Mesocricetus auratus* (hamster) | -49.7 | -50.0 | Permissive, high infectivity, lung disease [29,30] | -50.4 | |
| *Mustela putorius furo* (ferret) | -48.6 | -49.2 | Permissive, moderate infectivity, no overt disease [30-32] | -45.4 | |
| *Felis catus* (cat) | -47.6 | -48.9 | Permissive, high infectivity, lung disease [28,31,33] | -48.9 | predicted |
| *Panthera tigris* (tiger) | -47.3 | -42.5 | Permissive, overt disease, RNA positive[28] | -50.6 | |
| *Rhinolophus sinicus* (bat) | -46.9 | -50.1 ± 1.0 | Permissive? No reported infections [11] | -50.4 | confirmed |
| *Paguma larvata* (civet) | -45.1 | -46.1 | No reported infection | -49.4 | confirmed |
| *Equus ferus caballus* (horse) | -44.1 | -49.2 | No naturally occurring infections[28] | -48.0 | predicted |
| *Bos taurus* (cattle) | -43.6 | -42.5 | No naturally occurring infections[28] | | predicted |
| *Ophiophagus hannah* (king cobra) | -39.5 | -40.7 ± 1.2 | No reported infection | -45.4 | |
| *Mus musculus* (mouse) | -38.8 | -39.4 | Resistant to infection[30] | -44.7 | confirmed not used |

Golden hamsters, cattle and cats were members of the medium risk group and dogs, horses and bats were in the low risk group. Pangolins, ferrets, mice, and minks were assigned to the very low risk group in their analyses. The susceptibility predictions from these studies also do not correlate well with observations, consist with the low correlation between binding energies and sequence noted above.

Two additional, recent studies have some relevant to our study. Spinello et al. published a detailed, microsecond molecular dynamics simulation of the key molecular interactions driving the higher affinity/specificity of SARS-CoV-2 toward ACE2 as compared to SARS-CoV.[34] However, this study did not compare the affinities of non-human ACE2 proteins for the SARS-Cov-2 spike protein. Subsequently, Wang and co-workers published a complementary MD study comparing the interactions of SARS-Cov-2 and SARS-Cov spike proteins with the human ACE2 protein with 200 ns simulations.[35] They analysed the receptor binding of SARS-CoV-2 in terms of key electrostatic interactions, networks of hydrogen-bonding and hydrophobic interactions in the RBD. Again, no other species were considered in the study.

The key interacting residues of ACE2 receptor and SARS-CoV-2 S protein (Table 1) were broadly consistent with these previous studies [34-36]. Conserved amino acids in ACE2 of most species were largely those interacting with the S protein. PHE28, ASN330, ASP355 and ARG357 were conserved in all the species and all interacted with the spike protein. TYR41, LYS353, ALA386 and ARG393 were conserved and interacting in all species except bat, mouse, ferret and pangolin. Residues interacting with the S protein in ACE2 from *Ophiophagus hannah* (King cobra) were least conserved, consistent with its low sequence similarity to human ACE2.

**Computed binding affinities suggest that SARS-Cov-2 S protein is optimized for human ACE2**

Identifying species permissive to SARS-CoV-2 is very important for identifying intermediate hosts and potential source of the original virus, and for developing infection models for testing of COVID-19 drugs and vaccines. SARS-CoV-2 traces back to the human, civet, and bat SARS CoV strains, all of which use the ACE2 protein of each relevant species for cellular entry.[37-39] The SARS-CoV-2 and SARS CoV sequences have only 72% identity in the RBD regions showing they are only distantly related, even though they bind the same ACE2 receptor. The closest coronavirus to SARS-CoV-2 identified to date is the bat CoV, RaTG13, but it has a different spike protein that lacks the polybasic furin cleavage site and uses a different RBD. The RBD of pangolin CoV is most similar to that of SARS-CoV-2, meaning that much of early interpretation of the ancestry of SARS-CoV-2 was based on this sequence analysis (Figure 1).

Our structure-based approach revealed some surprisingly results, contrasting with to those from sequence based analyses. Conspicuously, the predicted binding between SARS-CoV-2 S protein and ACE2 was strongest overall for humans than for any species studied. The predicted strength of SARS-CoV-2 S protein binding to ACE2 being human > pangolin > dog > monkey > hamster > ferret > cat> tiger > bat > civet > horse > cow > snake > mouse ACE2.

Spike protein mutations are rare, especially in the receptor binding domain (RBD) that interacts with ACE2. For the SARS-Cov-2 S protein, only three mutational sites, V367F, G476S, and V483A, have been identified within the RBD domain. Of these, only G476S occurs directly at the binding interface of RBD and the peptidase domain of ACE2 and its incidence and geographic spread is very small.[40] This minimal mutation in the spike RBD during the history of the pandemic suggests that the spike RBD was already optimally adapted to bind human ACE2. This finding is surprising as, typically, a virus exhibits the highest affinity for receptors in its original host species, presumed by most to be bat, with a lower initial affinity for the receptor of any new host species such as humans. Then as the virus mutates and adapts to its new host species, the binding affinity might be expected to increase over time. Since our calculations are based on SARS-Cov-2 S sequences samples isolated in December and hence from very early in the outbreak, the finding that the affinity of SARS-CoV-2 S protein is higher for human ACE2 than for any other putative host species was unexpected.

This very high affinity for human ACE2 was confirmed very recently (July 10) in a preprint by Alexander et al.[41] who also studied the RBD of spike-ACE2 for several species and reported that the SARS-CoV-2 RBD sequence is optimal for binding to human ACE2 compared to other species. They also described this as a remarkable finding that underlies the high transmissibility of the SAR-Cov-2 virus amongst humans. These results are also consistent with a recent report comparing SARS-CoV and SARS-CoV-2 that found a number of differences in the SARS-CoV-2 RBD that made it a much more potent binder to human ACE2 through the introduction of numerous hydrogen bonding and hydrophobic networks[35] helping explain the efficient and rapid transmission SARS-CoV-2 through the human population, once a presumed cross-over event occurred in or around November 2019.

Interestingly, as shown in Table 1, pangolin and human ACE2 are closest in binding energy despite being structurally different and only sharing 10 of 16 interacting residues at the SARS-CoV-2 RBD. This similarity of binding energy is interesting as pangolins have previously been imputed as a potential intermediate host to explain the spill-over of a putative bat coronavirus to humans. However, most of the pangolin ACE2 RBD differences are conservative replacements of residues in the human ACE2 RBD viz. Q24E, D30E, D38E, L79I that are likely to make similar contributions to the binding interaction with SARS-CoV-2 spike. The main differences are the lack of S19 interaction and the replacement of H34 by S34 in pangolin ACE2. However, the MD structures show that the OH moiety in the sidechain of S24 lies in the same region, and can make similar interactions, as the NH moiety in the imidazole sidechain of H34.

**Affinity of ACE2 from non-human species**

SARS-CoV-2 also recognizes ACE2 from a variety of animal species, including palm civet, the intermediate host for SARS-CoV.[42] The calculated binding energies of SAR-CoV-2 S protein with ACE2 from other species is in general agreement with the relatively limited amount of published information on species infectivity. Predicted binding to monkey ACE2 is 2.4 kcal/mol lower than for human ACE2, despite all 16 ACE2 binding residues being the same This relatively small binding energy difference appears to reside in structural changes outside the RBD. Several studies have reported the susceptibility of various animal species to SARS-CoV-2[5,43,44] with susceptible species such as macaques, hamsters and ferrets used as animal models of SARS-CoV-2 infection.[29,45,46]

Infected young cynomolgus macaques expressed viral RNA in nasal swabs but did not develop overt clinical symptoms, whereas aged animals showed higher viral RNA loads, some weight loss, and rapid respiration associated with moderate interstitial pneumonia and virus replication in upper and lower respiratory tract.[46,47] Hamster ACE2 had a high predicted binding to the S protein, similar to that for monkey ACE2, with hamster ACE2 sharing 14 out of 16 binding residues with human ACE2. The model predicts that hamster should be permissive to SARS-CoV-2 infection based on the strong binding of SARS-Cov-2 S protein for hamster ACE2 (Table 2). Syrian hamsters have been shown to exhibit clinical and histopathological responses to SARS-CoV-2 that closely mimic human upper and lower respiratory tract infections, with high virus shedding and ability to transmit to naïve contact animals, [29] consistent with the model predictions. This high susceptibility to SARS-CoV infection of Syrian hamsters makes them one of the best animal models of infection. Ferrets are also permissive to SARS CoV infection, and our modelling data indicated that SARS-CoV-2 has a similar binding energy to ferret and hamster ACE2 (Table 2). Consistent with our model data, ferrets are permissive to infection with SARS-CoV-2, with high virus titre in the upper respiratory tract, virus shedding, infected ferrets showing acute bronchiolitis although without severe disease or death, and active transmission to naïve ferrets through direct contact.[44,45]

There is inefficient virus replication of SARS-CoV or SARS-CoV-2 in mice making them unsuitable as models to test SARS or COVID-19 vaccines or drugs. [48] Mice only became permissive for SARS infection when made transgenic for human ACE2 e.g. Golden et al. recently reported that transgenic mice expressing human ACE2 become highly susceptible to SARS-CoV-2.[49] Cat and tiger ACE2 were shown by our model to have similar binding affinity for SARS-CoV-2 S protein and both these species have been shown to be permissive for SARS-CoV-2 infection.

Similarly, our data suggests that SARS-CoV-2 binds with relatively high affinity to dog ACE2 suggesting that dogs may be susceptible to infection. In the case of companion animals that live in close proximity to humans, Shen at al. found that that SARS-CoV-2 could be efficiently transmitted in cats and dogs [50] while Shi et al. reported that SARS-CoV-2 replicates poorly in dogs, pigs, chickens, and ducks, but that ferrets and cats were permissive to infection.[31] Temmam et al. tested 9 cats and 12 dogs living in close contact with their owners, two of whom tested positive for SARS-CoV-2 and 11 of 18 others showed clinical signs of COVID-19 but no antibodies against SARS-CoV-2 were detectable in their blood using an immunoprecipation assay.[51] Goumeniu et al. published an editorial querying the role of dogs in the Lombardy COVID-19 outbreak and recommended use of computational docking experiments to provide evidence for, or against infection of dogs.[52] Our model data suggesting that dog ACE2 might be permissive for SARS-CoV-2 binding and infection is therefore consistent with anecdotal reports of dogs being infected with SARS-CoV-2. Genetic and other factors could underlie the apparent inconsistent reporting of susceptibility of dogs to COVID-19 clinical infection.

**Implications for the intermediate animal vector for SARS-Cov-2**

Bats have been suggested as the original host species of SARS-Cov-2 infections in humans, with pangolins acting as an intermediate animal vector. Bat CoV RaTG13 has the highest sequence similarity to SARS-CoV-2, with 96% whole-genome identity (50), but does not possess neither the furin cleavage site or the pangolin RBD seen in SARS-CoV-2. Could SARS-CoV-2 be an as-yet unidentified bat virus? Although bats carry many coronaviruses, no evidence of a direct relative of SARS-CoV-2 in bat populations has so far been found. As highlighted by our data, the binding affinity of SARS-CoV-2 for bat ACE2 is considerably lower than for human ACE2 and human ACE2 has been shown to not to bind the RaTG13 spike [53]. This suggests that even if SARS-CoV-2 did originally arise from a bat precursor, it must have spent considerable time in another host wherein it adapted its S protein to bind better to the host ACE2. This also resulted in acquiring higher affinity for human ACE2 while lowering its affinity to the original bat ACE2. There are currently no known explanations for how or where this transition to generate the ultimate SARS-CoV-2 spike protein could have occurred. Evidence of direct human infection by bat coronaviruses or other viruses is rare, with transmission typically involving an intermediate host. For example, SARS-CoV was shown to be transmitted from bats to civet cats and then to humans. The S protein must acquire new mutations to enable this transition to occur, first to increase its affinity for civet ACE2 and then to increase its affinity for human ACE2. To date, a virus directly related to SARS-CoV-2 has not been identified in bats or any other non-human species, making its origins unclear. Moreover, Wrobel et al. reported a structural biology study of the bat virus and SARS-Cov-2 and reported the experimental binding affinities of human ACE2 for the S proteins of these viruses.[53] They concluded that, although the structures of human SARS-CoV-2 and bat RaTG13 S proteins are similar, the human S protein has a more stable pre-cleavage form, and the human ACE2 binding $k_D$ was 68±9 nM while RaTG13 bound human ACE2 with a $k_D$ >40 µM. They observed that cleavage at the furin-cleavage site decreases the overall stability of SARS-CoV-2 S protein and fosters the open conformation required for S to bind to the ACE2 receptor. They concluded that their data indicate that a bat virus would not bind effectively to human ACE2 receptor and would be unlikely to infect humans. Early in the COVID-19 outbreak it was suggested that snakes may also be an intermediate vector. ACE2 of turtle and snake has very low homology to human ACE2 and SARS-CoV-2 was shown not to bind to reptile ACE2, making snakes unlikely intermediate hosts for SARS-CoV-2. [14]

The pangolin ACE2, despite having lower sequence similarity to human ACE2 (86%) than other tested species (e,g. monkey ACE2 - 97% similarity to human), had the most similar binding affinity to that of human ACE2. However, the predicted binding strength of SARS-CoV-2 S protein to pangolin ACE2 was lower than to human ACE2, with the difference being highly significant (two-tailed p = 0.0013). All other binding affinities were substantially lower then human and pangolin ACE2 and were significant at the >>99.99% confidence level. It was reported recently that a coronavirus isolated from Malayan pangolins shared the same RDB sequence as SARS-CoV-2, which raised the possibility that pangolins may have acted as the intermediate vector between bats and humans. Although it has the same spike RBD as SARS-CoV-2, pangolin-CoV is not closely related to SARS-CoV-2 overall, with <90% sequence similarity across its whole genome. [25] It is noteworthy that the common spike RBD in both pangolin CoV and SARS-CoV-2 is able to bind strongly to both pangolin and human ACE2, despite significant differences in the sequences of the ACE2 RDB (only 63% of residues are common (Table 1)). Xiao reported a pangolin-CoV that has 100%, 98.6%, 97.8% and 90.7% amino acid identity with SARS-CoV-2 in the E, M, N and S proteins.[26] They surprisingly reported the RBD of the pangolin S protein to be almost identical to that of SARS-CoV-2 despite 4 of the 10 residues interacting with ACE2 being different viz. H34S, L79I, M82N and G354H. They suggested pangolins as potential intermediate hosts of SARS-CoV-2. Additionally, pangolin CoV isolated from Malayan pangolins from two different regions in China showed differences in the residues interacting with human ACE2.[25] Pangolin-CoV was detected in 17 out of the 25 infected Malayan pangolins that that showed clinical signs and histological changes, and had antibodies against pangolin-CoV that also recognised the S protein of SARS-CoV-2. Another origin possibility might be that a pangolin was simultaneously co-infected with a bat ancestor to SARS-CoV-2 and a pangolin CoV. This could allow a recombination event to occur whereby the spike RBD of the pangolin virus was inserted into the bat CoV, thereby conferring the bat CoV with high binding for both pangolin and human ACE2. Such recombination events are known to occur with other RNA viruses and can explain creation of some pandemic influenza strains[54]. Nevertheless, such events are by necessity rare as they require coinfection of the one host at exactly the same time. Most importantly, if such a recombination event had occurred in pangolins it would be expected to have similarly triggered an epidemic spread of the new highly permissive SARS-CoV-2-like virus among pangolin populations, such as we now see occurring across the human population. Currently there is no evidence of such a widespread pangolin SARS-CoV-2 like outbreak, making this scenario less likely. Of course, some pangolins might have been protected from SARS-CoV-2 infection by the existence of neutralising antibodies cross-protective against both pangolin CoV and SARS-CoV-2, given their close RBD similarity. Notably, all pangolin coronaviruses identified so far lack the furin-like S1/S2 cleavage site in the SARS-CoV-2 spike protein that facilitates its rapid spread through human populations.

The high affinity of SARS-CoV-2 S protein for human ACE2 may be explicable if a recombination event took place in the past between a SARS-CoV-2 progenitor virus and pangolin CoV that resulted in insertion of the pangolin CoV RBD into the SARS-CoV-2 genome. However, the SARS-CoV-2 genome was recently reported to exhibit no evidence of recent recombination, arguing against this possibility.[55] Also, as he pangolin-CoV S protein does not have the furin cleavage site that is a prominent feature of the SARS-CoV-2 S protein,[51] this argues against pangolins being the intermediate vector for transmission of SARS-CoV-2 to humans. The major similarity of SARS-CoV-2 to pangolin-CoV lies in the RBD residues in the S protein that SARS-CoV-2 acquired by some unknown mechanism. As coronaviruses from nonhuman species with similar ACE2 receptors have not been sampled extensively, it is possible that a direct or intermediate animal source has yet to be found. This would preclude identification of evolutionary changes immediately prior to or during the jump to humans.

Gain of function (GoF) mutations occur in viruses that can lead to pandemics. For example, GoF mutations in influenza virus are associated with mammalian transmissibility, increased virulence for humans, and evasion of existing host immunity.[54] The conditioning of viruses to humans as pandemics progress is well recognized. Phylodynamic analyses of the COVID-19 genomes estimated the date for the most recent common ancestor from the start to middle of December, consistent with the earliest reported date 1st December 2019 for the initial cluster of pneumonia cases.[56] This study concluded, based on available genome sequence data, that the current epidemic has been driven entirely by human to human transmission at least since December. As the SARS-CoV-2 structure that we employed was obtained from viruses collected early in the outbreak, it is not clear how these early strains of SARS-CoV-2 developed such a high affinity for human ACE2. This suggests that SARS-CoV-2 spike RBD previously evolved by selection on a human-like ACE2. Notably, pangolin ACE2 bears major differences in its RBD to human ACE2. Therefore its is surprising that pangolin-CoV has a similar RBD to SARS-CoV-2 that forms the basis for implicating pangolins directly or indirectly in the origins of SARS-CoV-2. The fact that pangolin CoVs can use human ACE2 for cell entry suggests that pangolin CoVs could represent a source of future human coronavirus pandemics if they were to gain the SARS-CoV-2 furin cleavage site.

Given the seriousness of the ongoing SARS-CoV-2 pandemic, it is imperative that all efforts be made to identify the original source of the virus. One question to be addressed is whether the virus is completely natural and was transmitted to humans by an intermediate animal vector, or whether it came from a recombination event that occurred inadvertently or intentionally in a laboratory handling coronaviruses, with the new virus being inadvertently released into the local human population. This is of key importance given the ability to use such information to help prevent any similar outbreak in the future. The one positive to our observation that SARS-CoV-2 already has optimised high binding to human ACE2, providing little selective pressure for mutations in the spike RBD to further increase binding affinity. Thus, a vaccine that induces neutralising antibody to the spike RBD may remain effective long-term, and not need to be modified regularly like influenza vaccines to keep up with RBD mutations.

## Materials and Methods

**Homology modelling of S protein and ACE2 from multiple species.** As no three-dimensional structure of the SARS-CoV-2 S protein was available at the commencement of the project, we generated a homology structure the sequence retrieved from NCBI Genbank Database (accession number YP_009724390.1) in January 2020. A PSI-BLAST search against the PDB database for template selection was performed and the x-ray structure of SARS coronavirus S template (PDB ID 5XLR) was selected with 76.4% sequence similarity to SARS-CoV-2 S protein. The sequence alignment and sequences of related bat and pangolin coronaviruses are shown in Figure 1.

The protein sequences of the ACE2 proteins for different species and full sequence alignment in are shown in Supplementary Figure 2. The phylogenetic tree for ACE2 proteins from selected animal species is illustrated in Supplementary Figure 1. Protein preparation and removal of non-essential and non-bridging water molecules for docking studies and analysis of docked proteins was performed using the UCSF Chimera package (https://www.cgl.ucsf.edu/chimera/).[57]

The 3D-structures of the RBD of SARS-Cov-2 S and non-human ACE2 proteins were built using Modeller 9.23 (https://salilab.org/modeller/)[58] using the SARS S protein template to generate the homology model of SARS-Cov-2 S. The ACE2 receptors of selected species were similarly homology modelled using the following template structures – 1R42 (human ACE2), 3CSI (human glutathione transferase) and 3D0G (ACE2 structure from spike protein receptor-binding domain from the 2002-2003 SARS coronavirus human strain complexed with human-civet chimeric receptor ACE2) (Supplementary Table 1). Template similarity is important for model building the model; the sequence of *Macaca fascicularis* (monkey, accession number A0A2K5X283) was 97% similar to that of human ACE2 while *Ophiophagus hannah* (king cobra) had a much lower similarity of 61% this template. The quality of the generated models was evaluated using the GA341 score [59] and DOPE ((Discrete Optimized Protein Energy) method scores[60], and the models assessed using SWISS-MODEL structure assessment server (https://swissmodel.expasy.org/assess).[61] Structures with the lowest DOPE score were refined by MD simulations (*vide infra*) and used for further analysis.

The modelled structures were also assessed for quality control using Ramachandran Plot and molprobity scores in SWISSModel. The Ramachandran plot checks the stereochemical quality of a protein by analysing residue-by-residue geometry and overall structure geometry and visualizing energetically allowed regions for backbone dihedral angles ψ against φ of amino acid residues in protein structure. The Ramachandran score of SARS-CoV-2 spike protein was 90% in the binding region and the molprobity score was 3.17. The Ramachandran score of the percentage of amino acid residues in the various species ACE2 that fall into the energetically favoured region ranged from 96-99% (Supplementary Table 2). The predicted ACE2 structures and Ramachandran plots for each species are summarized in Supplementary Figure 3. The molprobity provides evaluation of model quality at both the global and local level, this combines protein quality score that reflects the crystallographic resolution of a model.[62] It is a log-weighted combination of the number of serious atom clashes per 1000 atoms, percentage Ramachandran not favoured, and percentage bad side-chain rotamers. A good molprobity score is one that is equal to or lower than the crystallographic resolution. For reference, the MolProbity score for the x-ray structures of the templates (PDB IDs) were: 1R42= 3.01; 3SCI=3.14; 3D0G=2.74 and 6M17=1.99. The Ramachandran and Molprobity scores show that all the built structures were of good quality and suitable for use in further studies.

The structure of the SARS-Cov-2 S protein was published subsequently (e.g. PDB ID 6VXX and 6VYB). [63] Figure 2 shows the very high structural similarity of our homology modelled spike protein structure with the EM structures (PDB ID 6LZG (RBD) and 6VYB (open state)) with RMSD of 0.36 Å.

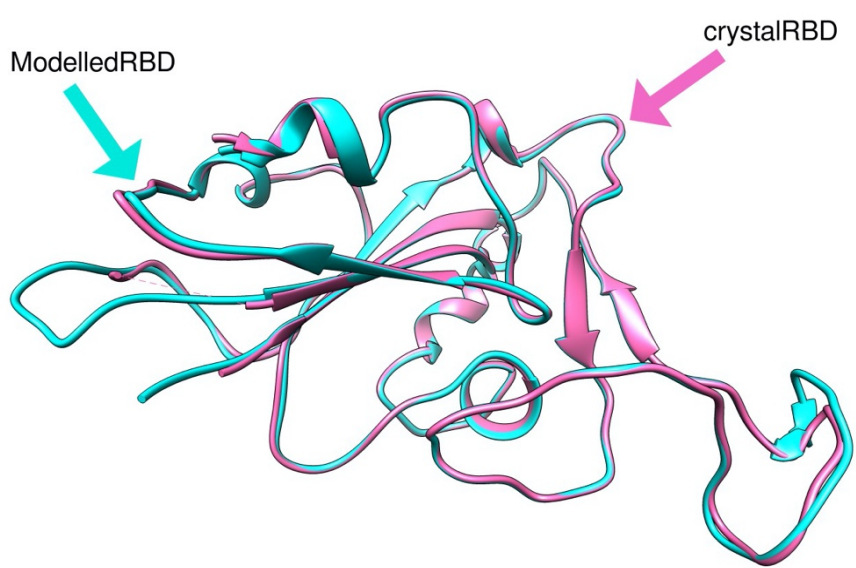

**Figure 2**. 3D structure of SARS-CoV-2 S protein (PDB PDB ID 6VYB (open form) and the homology modelled structure from Modeller

**Docking of SARS-Cov-2 S protein with ACE2 proteins**. These homology modelled ACE2 structures were docked against SARS-CoV-2 S protein structure using a state-of-the-art package HDOCK (http://hdock.phys.hust.edu.cn/).[64,65] This performs rigid-body docking by mapping the receptor and ligand molecules onto grids. Molecular docking was performed on the homology modelled SARS-CoV-2 S protein and human and animal ACE2 proteins using the hybrid docking method because attempts to use the template-free docking method with the structures generated by Modeller were unsatisfactory. The hybrid method used the structural template for the complex (PDB ID 6M17) to generate the results reported here. All docking poses were ranked using an energy-based scoring function.

The hybrid docking procedure may potentially introduce bias into the structures of the S protein bound to ACE2 from non-human species because it uses a human ACE2 complex x-ray structure as a template. To check for possible bias, the ACE2 structures generated by HDOCK were compared with those generated independently by Modeller. The Cα backbones of the ACE2 structures aligned with RMSD values between 0.5-0.8 Å (see Supplementary Table 3), exhibiting very strong structural similarities. Additionally, complexes were subjected to molecular dynamics simulation to wash out any template-induced bias.

**Molecular dynamics simulation of docked complexes**. The final docked SARS-Cov-2 spike/ACE2 protein complexes were optimized using the AMBER99SB-ILDN force field in GROMACS2020 (http://www.gromacs.org/).[66] Simulations were carried out using the GPU accelerated version of the program with the AMBER99SB-ILDN force field and implementing periodic boundary conditions in ORACLE server. Docked complexes were immersed in a truncated octahedron box of TIP3P water molecules. The solvated box was further neutralized with Na+ or Cl− counter ions using the tleap program. Particle Mesh Ewald (PME) was employed to calculate the long-range electrostatic interactions. The cut-off distance for the long-range van der Waals (VDW) energy term was 12.0 Å. The system was minimized without restraints. We applied 2500 cycles of steepest descent minimization followed by 5000 cycles of conjugate gradient minimization. After system optimization, the MD simulations was initiated by gradually heating each system in the NVT ensemble from 0 to 300 K for 50 ps using a Langevin thermostat with a coupling coefficient of 1.0/ps and with a force constant of 2.0 kcal/mol·Å2 on the complex. Finally, a production run of 100 ns of MD simulation was performed under a constant temperature of 300 K in the NPT ensemble with periodic boundary conditions for each system. During the MD procedure, the SHAKE algorithm was applied to all covalent bonds involving hydrogen atoms. The time step was 2 fs. The structural stability of the complex was monitored by the RMSD and RMSF values of the backbone atoms of the entire protein. Finally, the free energies of binding were calculated for all simulated docked structures.

Calculations were also performed for up to 500 ns on human ACE2 to ensure that 100ns is sufficiently long for convergence. Multiple production runs with different starting random seeds were used to estimate binding energy uncertainties for the strongest binding ACE2 structures – human, bat, and pangolin. All complexes stabilized during simulation with RMSD fluctuations converging to a range of 0.5 to 0.8 nm. The RMSD values for superimposition of the Cα backbones of each ACE2 structure before and after 100ns simulation was 1.2±0.1 Å, showing significant movement away from the initial HDOCK structures.

We found that the complex had stabilised after 50ns so considered that 100ns was an adequate simulation time. We also compared 100ns simulated structures of the ACE2 proteins from all species against those generated by homology modelling (most x-ray structures were not available) and found the RMSD values for Cα alignments between 0.5-0.8 Å. This suggests that any memory of the human template has been removed or minimized. We also compared the structures generated independently by homology (Modeller) and HDOCK (Supplementary Table 3) and they agreed very well (RMSD< 1Å)

**Calculation of binding free energies of complexes**. The binding free energies of the protein-protein complexes were evaluated in two ways. The traditional method is to calculate the energies of solvated SARS-CoV-2 S and ACE2 proteins and that of the bound complex proteins and derive the binding energy by subtraction.

$$\Delta G \text{ (binding, aq)} = G \text{ (complex, aq)} - G \text{ (spike, aq)} - G \text{ (ACE2, aq)} \quad (1)$$

We also calculated binding energies using the molecular mechanics Poisson Boltzmann surface area (MM-PBSA) tool in GROMACS that is derived from the nonbonded interaction energies of the complex.[67,68] The method is also widely used method for binding free energy calculations. The binding free energies of the protein complexes were analysed during equilibrium phase from the output files of 100 ns MD simulations. The `g_mmpbsa` tool in GROMACS was used after molecular dynamics simulations, the output files obtained were used to post-process binding free energies by the single-trajectory MM-PBSA method. Specifically, for a non-covalent binding interaction in the aqueous phase the binding free energy,

$$\Delta G \text{ (bind,aqu)} = \Delta G \text{ (bind,vac)} + \Delta G \text{ (bind,solv)} \quad (2)$$

where $\Delta G \text{ (bind,vac)}$ is the binding free energy in vacuum, and $\Delta G\text{(bind,solv)}$ is the solvation free energy change upon binding: –

$$\Delta G \text{ (bind,solv)} = G \text{ (R:L, solv)} - G \text{ (R,solv)} - G \text{ (L,solv)} \quad (3)$$

where G (R:L,solv), G (R,solv) and G (L,solv) are solvation free energies of complex, receptor and ligand, respectively.[69]

Free energy decomposition analyses were also performed by MM-PBSA decomposition to get a detailed insight into the interactions between the ligand and each residue in the binding site. The binding interaction of each ligand–residue pair includes three terms: the van der Waals contribution, the electrostatic contribution, and the solvation contribution.

As the simulations are very lengthy, we only ran multiple simulations for the species proposed as intermediate hosts and potential sources of the original virus, human, pangolin, bat and snake to estimate the uncertainty in the binding energies. As the ACE2 proteins for all species were extremely similar, we expected that these simulation error estimates would be of the same order for all other species. The predicted binding energies for these other species were significantly lower than those of human ACE2. We also used a statistical test to calculate the probability of the pangolin and human ACE2 affinities being different.

**Data Availability**. The coordinates of the S protein-ACE2 complexes will be deposited in. data repositories at La Trobe University and Flinders University.

**ACKNOWLEDGEMENTS**

We would like to thank Harinda Rajapaksha for assistance to optimise GROMACS for this project. We would also like to thank Oracle for providing their Cloud computing resources for the modelling studies described herein. In particular, we wish to thank Peter Winn, Dennis Ward, and Alison Derbenwick Miller from Oracle in facilitating these studies. The opinions expressed herein are solely those of the individual authors and should not be inferred to reflect the views of their affiliated institutions, funding bodies or Oracle corporation.

**Author contributions**

Petrovsky - conceived project, analysed data, contributed to manuscript; Piplani and Singh - performed the computations, analysed data, contributed to the manuscript; Winkler - analysed data and contributed to manuscript

# In silico comparison of SARS-CoV-2 spike protein-ACE2 binding affinities across species and implications for viral origin

Sakshi Piplani[1,2], Puneet Kumar Singh[2], David A. Winkler[3-6]*, Nikolai Petrovsky[1,2]*

[1] College of Medicine and Public Health, Flinders University, Bedford Park 5046, Australia

[2] Vaxine Pty Ltd, 11 Walkley Avenue, Warradale 5046, Australia

[3] La Trobe University, Kingsbury Drive, Bundoora 3086, Australia

[4] Monash Institute of Pharmaceutical Sciences, Monash University, Parkville 3052, Australia

[5] School of Pharmacy, University of Nottingham, Nottingham NG7 2RD. UK

[6] CSIRO Data61, Pullenvale 4069, Australia

**SUPPLEMENTARY INFORMATION**

**Supplementary Table 1**. Human template structures used to model selected ACE2 species and similarity scores of each ACE2 sequence to the selected template used.

| **Species** | **Accession No.** | **Database** | **Template** | **Similarity** |
|---|---|---|---|---|
| *Rhinolophus sinicus* (Bat) | AGZ48803.1 | UniProt | 3SCI | 79.73% |
| *Mus musculus* (Mouse) | Q8R0I0 | UniProt | 1R42 | 84.27% |
| *Mustela putorius furo* (Ferret) | Q2WG88 | UniProt | 1R42 | 83.44% |
| *Mesocricetus auratus* (Hamster) | A0A1U7QTA1 | UniProt | 1R42 | 87.58% |
| *Felis catus* (Cat) | Q56H28 | UniProt | 1R42 | 85.93% |
| *Canis luparis* (Dog) | J9P7Y2 | UniProt | 1R42 | 84.93% |
| *Paguma larvata* (Ccivet) | Q56NL1 | UniProt | 3D0G† | 86.77% |
| *Macaca fascicularis* (Monkey) | A0A2K5X283 | UniProt | 1R42 | 96.91% |
| *Manis javanica* (Pangolin) | XP_017505752.1 | NCBI | 1R42 | 85.57% |
| *Ophiophagus hannah* (King cobra) | V8NIH2 | UniProt | 1R42 | 61.42% |
| *Equus caballus* (Horse) | F6V9L3 | UniProt | 6MI7 | 85.91% |
| *Panthera tigris altaica* (Tiger) | XP_007090142.1 | NCBI | 6MI7 | 85.91% |
| *Bos taurus* (Cow) | NP_001019673.2 | NCBI | 6MI7 | 80.30% |

† ACE2 structure from spike protein receptor-binding domain from the 2002-2003 SARS coronavirus human strain complexed with human-civet chimeric receptor ACE2

**Supplementary Table 2.** MolProbity and Ramachandran scores for ACE2 modelled structures for selected species

| Species | MolProbity Score | Ramachandran Score (favoured region) |
|---|---|---|
| *Rhinolophus sinicus* (bat) | 2.45 | 97.8% |
| *Mus musculus* (mouse) | 2.49 | 98.2% |
| *Mustela putorius furo* (ferret) | 2.5 | 98.3% |
| *Mesocricetus auratus* (hamster) | 2.59 | 98.7% |
| *Felis catus* (cat) | 2.42 | 98.5% |
| *Paguma larvata* (civet) | 2.52 | 97.3% |
| *Macaca fascicularis* (monkey) | 2.59 | 98.2% |
| *Manis javanica* (pangolin) | 2.42 | 98.3% |
| *Ophiophagus hannah* (king cobra) | 2.93 | 96.0% |
| *Canis lupus familiaris* (dog) | 2.9 | 97.1% |
| *Equus caballus* (horse) | 2.29 | 96.3% |
| *Panthera tigris altaica* (tiger) | 2.25 | 96.6% |
| *Bos Taurus* (cattle) | 2.70 | 96.2% |

**Supplementary Table 3**. RMSD for alignments of Cα backbones for ACE2, and the whole ACE2spike complex generated by homology modelling (Modeller) and HDOCK.

| Species | RMSD Å Cα ACE2 | RMSD Å complex |
|---|---|---|
| Bat | 0.82 | 0.88 |
| Cat | 0.54 | 0.85 |
| Cattle | 0.77 | 0.79 |
| Civet | 0.74 | 0.85 |
| Dog | 0.69 | 0.82 |
| Ferret | 0.59 | 0.84 |
| Hamster | 0.60 | 0.85 |
| Horse | 0.82 | 0.84 |
| Monkey | 0.56 | 0.85 |
| Pangolin | 0.54 | 0.86 |
| Snake | 0.84 | 0.89 |
| Tiger | 0.87 | 0.85 |
| Mouse | 0.54 | 0.86 |

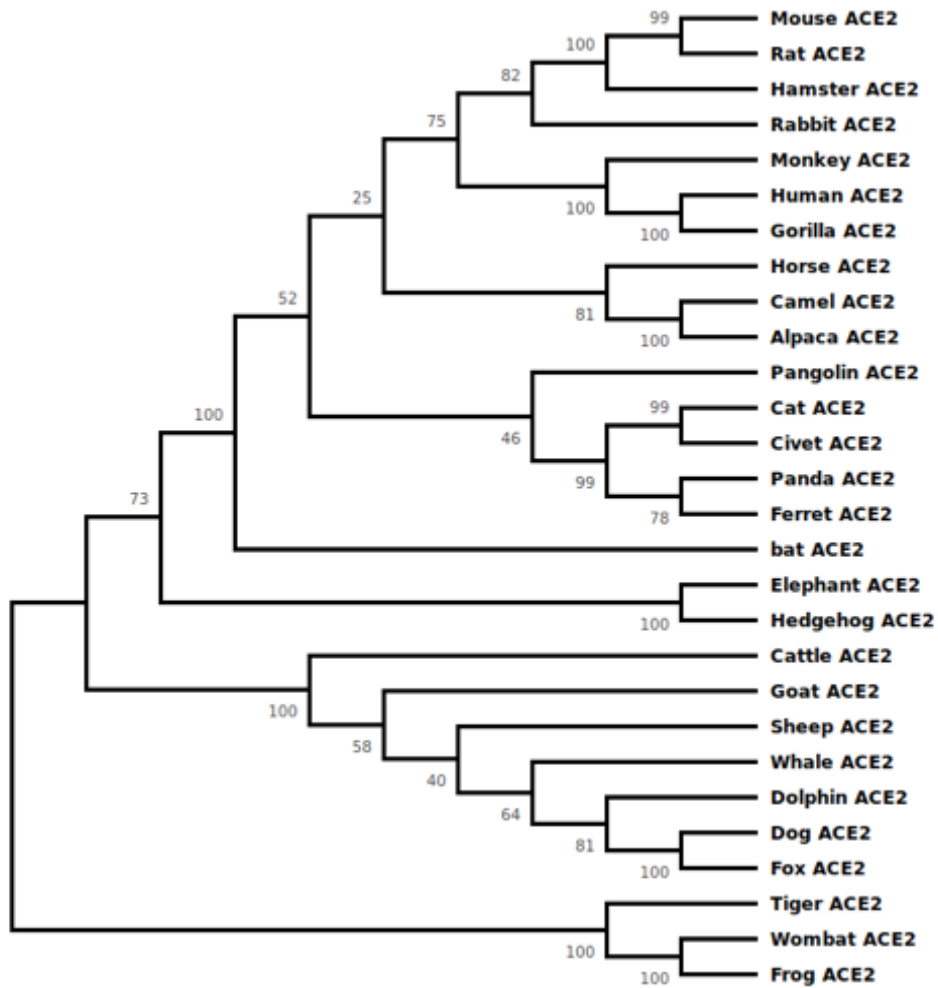

**Supplementary Figure 1**. Phylogenetic tree showing relatedness of sequences of ACE2 proteins from selected species.

```
Ophiophagus    MLMKQAPVRKPSSRSFTHPAFFDLKGNMLTWLCLTWSLVVLALAQ-DETKVATKFLEQFD   59
Mesocricetus   -------------------------MSSSSWLLLS--LVAVTTAQSIIEEQAKTFLDKFN   33
Mus            -------------------------MSSSSWLLLS--LVAVTTAQSLTEENAKTFLNNFN   33
Rhinolophus    -------------------------MSGSSWLLLS--LVAVTTAQSTTEDEAKMFLDKFN   33
Bos            -------------------------MTGSFWLLLS--LVAVTAAQSTTEEQAKTFLEKFN   33
Homo           -------------------------MSSSSWLLLS--LVAVTAAQSTIEEQAKTFLDKFN   33
Macaca         -------------------------MSGSSWLLLS--LVAVTAAQSTIEEQAKTFLDKFN   33
Mustela        -------------------------MLGSSWLLLS--LAALTAAQSTTEDLAKTFLEKFN   33
Canis          -------------------------MSGSSWLLLS--LAALTAAQS-TEDLVKTFLEKFN   32
Paguma         -------------------------MSGSFWLLLS--FAALTAAQSTTEELAKTFLETFN   33
Felis          -------------------------MSGSFWLLLS--FAALTAAQSTTEELAKTFLEKFN   33
Panthera       -------------------------------LS--FAALTAAQSTTEELAKTFLEKFN   25
Manis          -------------------------MSGSSWLLLS--LVAVTAAQSTSDEEAKTFLEKFN   33
Equus          -------------------------MSGSSWLLLS--LVAVTAAQSTTEDLAKTFLEKFN   33
                                                       *:   :..:: **      .   ..  **: *:

Ophiophagus    ARATDLYYNASIASWNYNTNLTEENAKIMHEKDNIFSKFYGEACRNASMFNVNHITDETI  119
Mesocricetus   QEAEDLSYQSALASWNYNTNITEENAQKMNEAAAKWSAFYEEQSKLAKNYSLQEVQNLTI   93
Mus            QEAEDLSYQSSLASWNYNTNITEENAQKMSEAAAKWSAFYEEQSKTAQSFSLQEIQTPII   93
Rhinolophus    TKAEDLSHQSSLASWDYNTNINDENVQKMDEAGAKWSAFYEEQSKLAKNYSLEQIQNVTV   93
Bos            HEAEDLSYQSSLASWNYNTNITEENVQKMNEARAKWSAFYEEQSRMAKTYSLEEIQNLTL   93
Homo           HEAEDLFYQSSLASWNYNTNITEENVQNMNNAGDKWSAFLKEQSTLAQMYPLQEIQNLTV   93
Macaca         HEAEDLFYQSSLASWNYNTNITEENVQNMNNAGEKWSAFLKEQSTLAQMYPLQEIQNLTV   93
Mustela        YEAEELSYQNSLASWNYNTNITDENIQKMNIAGAKWSAFYEEESQHAKTYPLEEIQDPII   93
Canis          YEAEELSYSWNYNTNITDENVQKMNNAGAKWSAFYEEQSKLAKTYPLEEIQDSTV   92
Paguma         YEAQELSYQSSVASWNYNTNITDENAKNMNEAGAKWSAYYEEQSKLAQTYPLAEIQDAKI   93
Felis          HEAEELSYQSSLASWNYNTNITDENVQKMNEAGAKWSAFYEEQSKLAKTYPLAEIHNTTV   93
Panthera       HEAEELSYQSSLASWNYNTNITDENVQKMNEAGAKWSAFYEEQSKLAETYPLAEIHNTTV   85
Manis          SEAEELSYQSSLASWNYNTNITDENVQKMNVAGAKWSTFYEEQSKIAKNYQLQNIQNDTI   93
Equus          SEAEELSHQSSLASWSYNTNITDENVQKMNEAGARWSAFYEEQCKLAKTYPLEEIQNLTV   93
                  .*   :*   ::   ::***.**  *:.:**   :  *          :*:     *     *.    :  .:        :

Ophiophagus    KLQIRLLQSGSTD----STKDQLDTVLHKMSTLYS-------------------LDDIMA  156
Mesocricetus   KRQLQALQQSGSSALSADKNKQLNTILNTMSTIYSTGKVCNPKNPQECLLLEPGLDDIMA  153
Mus            KRQLQALQQSGSSALSADKNKQLNTILNTMSTIYSTGKVCNPKNPQECLLLEPGLDEIMA  153
Rhinolophus    KLQLQILQQSGSPVLSEDKSKRLNSILNAMSTIYSTGKVCKPNKPQECLLLEPGLDNIMG  153
Bos            KRQLKALQHSGTSALSAEKSKRLNTILNKMSTIYSTGKVLDPN-TQECLALEPGLDDIME  152
Homo           KLQLQALQQNGSSVLSEDKSKRLNTILNTMSTIYSTGKVCNPDNPQECLLLEPGLNEIMA  153
Macaca         KLQLQALQQNGSSVLSEDKSKRLNTILNTMSTIYSTGKVCNPNNPQECLLLDPGLNEIME  153
Mustela        KRQLRALQQSGSSVLSADKRERLNTILNAMSTIYSTGKACNPNNPQECLLLEPGLDDIME  153
Canis          KRQLRALQHSGSSVLSADKNQRLNTILNSMSTIYSTGKACNPSNPQECLLLEPGLDDIME  152
Paguma         KRQLQALQQSGSSVLSADKSQRLNTILNAMSTIYSTGKACNPNNPQECLLLEPGLDNIME  153
Felis          KRQLQALQQSGSSVLSADKSQRLNTILNAMSTIYSTGKACNPNNPQECLLLEPGLDDIME  153
Panthera       KRQLQALQQSGSSVLSADKSQRLNTILNAMSTIYSTGKACNPNNPQECLLLEPGLDDIME  145
Manis          KRQLQALQLSGSSALSADKNQRLNTILNTMSTIYSTGKVCNPGNPQECSLLEPGLDNIME  153
Equus          KRQLQALQQSGSSVLSADKSKRLNEILNTMSTIYSTGKVCNPSNPQECLLLEPGLDAIME  153
                 * *::  **  ..:      ..  .:*:  :*:  ***:**                     *:  **

Ophiophagus    NNWNYPERLWAWEGWRANVGKKMRPLYETYVELKNKYARLRGYADYGDYWRANYEVDLPG  216
Mesocricetus   TSTDYNERLWAWEGWRAEVGKQLRPLYEEYVVLKNEMARANNYEDYGDYWRGDYEAEGAD  213
Mus            TSTDYNSRLWAWEGWRAEVGKQLRPLYEEYVVLKNEMARANNYNDYGDYWRGDYEAEGAD  213
Rhinolophus    TSKDYNERLWAWEGWRAEVGKQLRPLYEEYVVLKNEMARGYHYEDYGDYWRRDYETEESP  213
Bos            NSRDYNRRLWAWEGWRAEVGKQLRPLYEEYVVLENEMARANNYEDYGDYWRGDYEVTGAG  212
Homo           NSLDYNERLWAWESWRSEVGKQLRPLYEEYVVLKNEMARANHYEDYGDYWRGDYEVNGVD  213
Macaca         KSLDYNERLWAWEGWRSEVGKQLRPLYEEYVVLKNEMARANHYKDYGDYWRGNYEVNGVD  213
Mustela        NSKDYNERLWAWEGWRSEVGKQLRPLYEEYVALKNEMARANNYEDYGDYWRGDYEEEWAD  213
Canis          NSKDYNERLWAWEGWRSEVGKQLRPLYEEYVALKNEMARANNYEDYGDYWRGDYEEEWEN  212
Paguma         NSKDYNERLWAWEGWRAEVGKQLRPLYEEYVALKNEMARANNYEDYGDYWRGDYEEEWTG  213
Felis          NSKDYNERLWAWEGWRAEVGKQLRPLYEEYVALKNEMARANNYEDYGDYWRGDYEEEWTD  213
Panthera       NSKDYNERLWAWEGWRAEVGKQLRPLYEEYVALKNEMARANNYEDYGDYWRGDYEEEWTD  205
Manis          SSKDYNERLWAWEGWRSEVGKQLRPLYEEYVVLKNEMARANHYEDYGDYWRGDYEAEGAN  213
Equus          NSKDYNQRLWAWEGWRSEVGKQLRPLYEEYVVLKNEMARANNYEDYGDYWRGDYEAEGPS  213
                   .. :* ******.**::***::***** ** *:*: **     * ******* :**

Ophiophagus    KFQYQREQLITDVESTFKQ------QLHAYVRHHLYKRYGPELINPEGAIPAHLLGDMWG  270
Mesocricetus   GYNYNGNQLIEDVERTFKEIKPLYEQLHAYVRTKLMNTY-PSYISPTGCLPAHLLGDMWG  272
Mus            GYNYNRNQLIEDVERTFAEIKPLYEHLHAYVRRKLMDTY-PSYISPTGCLPAHLLGDMWG  272
Rhinolophus    GPGYSRDQLMKDVERIFTEIKPLYEHLHAYVRAKLMDTY-PFHISPTGCLPAHLLGDMWG  272
Bos            DYDYSRDQLMKDVERTFAEIKPLYEQLHAYVRAKLMHTY-PSYISPTGCLPAHLLGDMWG  271
Homo           GYDYSRGQLIEDVEHTFEEIKPLYEHLHAYVRAKLMNAY-PSYISPIGCLPAHLLGDMWG  272
Macaca         GYDYNRDQLIEDVERTFEEIKPLYEHLHAYVRAKLMNAY-PSYISPTGCLPAHLLGDMWG  272
Mustela        GYSYSRNQLIEDVEHTFQIKPLYEHLHAYVRAKLMDAY-PSRISPTGCLPAHLLGDMWG  272
Canis          GYNYSRNQLIDDVEHTFTQIMPLYEHLHAYVRTKLMDTY-PSYISPTGCLPAHLLGDMWG  271
Paguma         GYNYSRNQLIQDVEDTFEQIKPLYQHLHAYVRAKLMDTY-PSRISRTGCLPAHLLGDMWG  272
Felis          GYNYSRSQLIKDVEHTFTQIKPLYQHLHAYVRAKLMDTY-PSRISPTGCLPAHLLGDMWG  272
Panthera       GYNYSRSQLIKDVEHTFTQIKPLYQHLHAYVRAKLMDSY-PSRISPTGCLPAHLLGDMWG  264
Manis          GYNYSRDHLIEDVEHIFTQIKPLYEHLHAYVRAKLMDNY-PSHISPTGCLPAHLLGDMWG  272
Equus          GYDYSRDQLIEDVERTFAEIKPLYEHLHAYVRAKLMDTY-PSHINPTGCLPAHLLGDMWG  272
                    *.   :*: ***  *  :       :****** :* . * *   *.    *.:**********

Ophiophagus    RFWTNLYPLMVPYPNKTSIDVTSAMEKKKWTVNSIFKAAEHFFISIGLFNMTVGFWKNSM  330
Mesocricetus   RFWTNLYPLTVPFGQKPNIDVTDAMVNQGWNAERIFKEAEKFFVSVGLPYMTQGFWENSM  332
Mus            RFWTNLYPLTVPFAQKPNIDVTDAMMNQGWDAERIFQEAEKFFVSVGLPHMTQGFWANSM  332
Rhinolophus    RFWTNLYPLTVPFGQKPNIDVTDEMLKQGWDADRIFKEAEKFFVSVGLPNMTEGFWNNSM  332
Bos            RFWTNLYSLTVPFEHKPSIDVTEKMENQSWDAERIFKEAERIFKEMEKFFVSISLPYMTQGFWENSM  331
Homo           RFWTNLYPLTVPFGQKPNIDVTDAMVDQAWDAQRIFKEAEKFFVSVGLPNMTQGFWENSM  332
Macaca         RFWTNLYSLTVPFGQKPNIDVTDAMVNQAWNAQRIFKEAEKFFVSVGLPNMTQGFWENSM  332
Mustela        RFWTNLYPLMVPFRQKPNIDVTDGMVNQSWDARRIFEEAETFFVSVGLPNMTEGFWQNSM  332
Canis          RFWTNLYPLTVPFGQKPNIDVTNAMVNQSWDARKIFKEAEKFFVSVGLPNMTQEFWENSM  331
Paguma         RFWTNLYPLTVPFGQKPNIDVTDAMVNQNWDARRIFKEAEKFFVSVGLPNMTQGFWENSM  332
Felis          RFWTNLYPLTVPFGQKPNIDVTDAMVNQSWDARRIFKEAEKFFVSVGLPNMTQGFWENSM  332
Panthera       RFWTNLYPLTVPFGQKPNIDVTDAMVNQSWDARRIFKEAEKFFVSVGLPNMTQGFWENSM  324
Manis          RFWTNLYPLTVPFRQKPNIDVTDAMVNQTWDANRIFKEAEKFFVSVGLPKMTQTFWENSM  332
Equus          RFWTNLYSLTVPFGQKPNIDVTDAMVDQSWDAKRIFEEAEKFFVSVGLPNMTQGFWENSM  332
               ****** * **: :* .**** * * .: *  .   **: ** **:*:.*  **  ** ***

Ophiophagus    LEEPKGGRKVVCHPTAWDMGKEDYRIKMCTKINMEDFLTAHHEMGHIEYDMAYANQPFLL  390
Mesocricetus   LTDPGDDRKVVCHPTAWDLGKGDFRIKMCTKVTMDNFLTAHHEMGHIQYDMAYATQPFLL  392
Mus            LTEPADGRKVVCHPTAWDLGHGDFRIKMCTKVTMDNFLTAHHEMGHIQYDMAYARQPFLL  392
Rhinolophus    LTEPGDGRKVVCHPTAWDLGKGDFRIKMCTKVMEDFLTAHHEMGHIQYDMAYASQPYLL  392
Bos            LTEPGDGRKVVCHPTAWDLGKGDFRIKMCTKVTMDDFLTAHHEMGHIQYDMAYAAQPYLL  391
Homo           LTDPGNVQKAVCHPTAWDLGKGDFRILMCTKVTMDDFLTAHHEMGHIQYDMAYAAQPFLL  392
Macaca         LTDPGNVQKVVCHPTAWDLGKGDFRIIMCTKVTMDDFLTAHHEMGHIQYDMAYAAQPFLL  392
Mustela        LTEPGDNRKVVCHPTAWDLGKRDFRIKMCTKVTMDDFLTAHHEMGHIQYDMAYAEQPFLL  392
Canis          LTEPSDSRKVVCHPTAWDLGKGDFRIKMCTKVTMDDFLTAHHEMGHIQYDMAYAAQPFLL  391
Paguma         LTEPGDGRKVVCHPTAWDLGKGDFRIKMCTKVTMDDFLTAHHEMGHIQYDMAYAAQPFLL  392
Felis          LTEPGDSRKVVCHPTAWDLGKGDFRIKMCTKVTMDDFLTAHHEMGHIQYDMAYAVQPFLL  392
Panthera       LTEPGNSQKVVCHPTAWDLGKGDFRIKMCTKVTMDDFLTAHHEMGHIQYDMAYAVQPFLL  384
Manis          LTEPGDGRKVVCHPTAWDLGKHDFRIKMCTKVTMDDFLTAHHEMGHIQYDMAYAMQPYLL  392
Equus          LTEPGDGRKVVCHPTAWDLGKGDFRIKMCTKVTMDDFLTAHHEMGHIQYDMAYAVQPYLL  392
               * :*  .  :*.*********:*. *:** ****:.*::*********** ** **:**
```

```
Ophiophagus    RNGANEGFHEAVGEIMSLSAATPKYLKSLGLLEPTFQEDAETDINFLLKQALTIVGTMPF  450
Mesocricetus   RNGANEGFHEAVGEIMSLSAATPEHLKSIGLLPSDFQEDNETEINFLLKQALTIVGTLPF  452
Mus            RNGANEGFHEAVGEIMSLSAATPKHLKSIGLLPSDFQEDSETEINFLLKQALTIVGTLPF  452
Rhinolophus    RNGANEGFHEAVGEVMSLSVATPKHLKTMGLLSPDFREDNETEINFLLKQALNIVGTLPF  452
Bos            RNGANEGFHEAVGEIMSLSAATPHYLKALGLLAPDFHEDNETEINFLLKQALTIVGTLPF  451
Homo           RNGANEGFHEAVGEIMSLSAATPKHLKSIGLLSPDFQEDNETEINFLLKQALTIVGTLPF  452
Macaca         RNGANEGFHEAVGEIMSLSAATPKHLKSIGLLSPDFQEDNETEINFLLKQALTIVGTLPF  452
Mustela        RNGANEGFHEAVGEIMSLSAATPNHLKNIGLLPPDFSEDSETDINFLLKQALTIVGTLPF  452
Canis          RNGANEGFHEAVGEIMSLSAATPNHLKNIGLLPPSFFEDNETEINFLLKQALTIVGTLPF  451
Paguma         RNGANEGFHEAVGEIMSLSAATPNHLKTIGLLSPAFSEDNETEINFLLKQALTIVGTLPF  452
Felis          RNGANEGFHEAVGEIMSLSAATPNHLKTIGLLPPGFSEDSETEINFLLKQALTIVGTLPF  452
Panthera       RNGANEGFHEAVGEIMSLSAATPNHLKTIGLLPPGFSEDSETEINFLLKQALTIVGTLPF  444
Manis          RNGANEGFHEAVGEIMSLSAATPNHLKTIGLLPPDFYEDNETEINFLLKQALTIVGTLPF  452
Equus          RNGANEGFHEAVGEIMSLSAATPNHLKAIGLLPPDFYEDSETEINFLLKQALTIVGTLPF  452
               ****************.****.***.:.**.:***.   *.**.**.:*********..****.:.**

Ophiophagus    TYMLEKWRWMVFAEQIPKDQWMKKWWEMKREIVGVVEPLPHNEEYCDPAALFHVANDYSF  510
Mesocricetus   TYMLEKWRWMVFKGDIPKEQWMKEKWWEMKREIVGVVEPLPHDETYCDPAALFHVSNDYSF  512
Mus            TYMLEKWRWMVFRGEIPKEQWMKKWWEMKREIVGVVEPLPHDETYCDPASLFHVSNDYSF  512
Rhinolophus    TYMLEKWRWMVFKEEWMKKWWEMKRKIVGVVEPVPHDETYCDPSPDFREDPASLFHVANDYSF  512
Bos            TYMLEKWRWMVFKGEIPKQQWMEKWWEMKREIVGVVEPLPHDETYCDPACLFHVAEDYSF  511
Homo           TYMLEKWRWMVFKGEIPKDQWMKKWWEMKREIVGVVEPVPHDETYCDPASLFHVSNDYSF  512
Macaca         TYMLEKWRWMVFKGEIPKEQWMKKWWEMKREIVGVVEPLPHDETYCDPASLFHVANDYSF  512
Mustela        TYMLEKWRWMVFKGEIPKEQWMQKWWEMKRDIVGVVEPLPHDETYCDPAALFHVANDYSF  512
Canis          TYMLEKWRWMVFKGEIPKDQWMKTWWEMKRNIVGVVEPVPHDETYCDPASLFHVANDYSF  511
Paguma         TYMLEKWRWMVFKGAIPKEQWMQKWWEMKRNIVGVVEPVPHDETYCDPASLFHVANDYSF  512
Felis          TYMLEKWRWMVFKGEIPKEQWMKKWWEMKREIVGVVEPVPHDETYCDPASLFHVANDYSF  512
Panthera       TYMLEKWRWMVFKGEIPKEQWMQKWWEMKREIVGVVEPVPHDETYCDPASLFHVANDYSF  504
Manis          TYMLEKWRWMVFSGQIPKEQWMKKWWEMKREIVGVVEPVPHDETYCDPASLFHVANDYSF  512
Equus          TYMLEKWRWMVFKGEIPKEEWMKKWWEMKREIVGVVEPVPHDETYCDPAALFHVANDYSF  512
               ***********     ***: :**:.*****.*******.**:*  **** ****::***

Ophiophagus    IRYYTRTIYQFQFQEALCKAAGHTKELYKCDISDSTNAGRILKDMLALGSSQPWTKALES  570
Mesocricetus   IRYYTRTIYQFQFQEALCQAAKHDGPLHKCDISNSTEAGQKLLNMLRLGKSEPWTLALEN  572
Mus            IRYYTRTIYQFQFQEALCQAAKYNGSLHKCDISNSTEAGQKLLKMLSLGNSEPWTLALEN  572
Rhinolophus    IRYYTRTIFEFQFHEALCRIAQHDGPLHKCDISNSTDAGKKLHQMLSVGKSQAWTKTLED  572
Bos            IRYYTRTYQFQFQHEALCKTAKHEGALFKCDISNSTEAGQRLLQMLRLGKSEPWTLALEN  571
Homo           IRYYTRTLYQFQFQEALCQAAKHEGPLHKCDISNSTEAGQKLFNMLRLGKSEPWTLALEN  572
Macaca         IRYYTRTLYQFQFQEALCQAAKHEGPLHKCDISNSTEAGQKLLNMLKLGKSEPWTLALEN  572
Mustela        IRYYTRTIYQFQFQEALCQIAKHEGPLYKCDISNSSEAGQKLHEMLSLGRSKPWTFALER  572
Canis          IRYYTRTIYQFQFQEALCQIAKHEGPLHKCDISNSSEAGQKLLEMLKLGKSKPWTYALEI  571
Paguma         IRYYTRTIYQFQFQEALCQIAKHEGPLHKCDISNSTEAGKKLLEMLSLGRSEPWTLALER  572
Felis          IRYYTRTIYQFQFQEALCRIAKHEGPLHKCDISNSSEAGKKLQMLTLGSKPWTLALEH  572
Panthera       IRYYTRTIYQFQFQEALCRIAKHEGPLHKCDISNSSEAGKKLQMLTLGSKPWTLALEH  564
Manis          IRYYTRTIYQFQFQEALCQTAKHEGPLHKCDISNSAEAGQKLQMLSLGKSKPWTLALER  572
Equus          IRYYTRTIYQFQFQEALCQTAKHEGPLHKCDISNSTEAGQKLLQMLSLGKSEPWTLALER  572
               ********::::***:* :  *. :      .*.****::*::  .*  *  :*  :**

Ophiophagus    ITGSLKMDAKPFCQYFDPLLKWLEKTNSNENVGWNVNWTPYSKDAIKVRISLKAALGDDA  630
Mesocricetus   VVGARNMDVRPLLNYFEPLSVWLKEQNKNSFVGWNTDWSPYADQSIKVRISLKSALGENA  632
Mus            VVGARNMDVKPLLNYFQPLFDWLKEQNRNSFVGWNTEWSPYADQSIKVRISLKSALGANA  632
Rhinolophus    IVDSRNMDVGPLLKYFEPLYTWLQEQNRKSYVGWNTDWSPYSDQSIKVRISLKSALGENA  632
Bos            IVGIKTMDVKPLLNYFEPLFTWLKEQNRNSFVGWSTEWTPYSDQSIKVRISLKSALGENA  631
Homo           VVGAKNMNVRPLLNYFEPLFTWLKDQNKNSFVGWSTDWSPYADQSIKVRISLKSALGDKA  632
Macaca         VVGAKTMDVRPLLNYFEPLFTWLKDQNKNSFVGWSTDWSPYADQSIKVRISLKSALGDKA  632
Mustela        VVGAKNMDVRPLLNYFEPLFTWLKEQNRNSFVGWNTDWSPYADQSIKVRISLKSALGEKA  631
Canis          VVGAKNMNVTPLLNYFEPLFTWLKEQNRNSFVGWDTDWRPYSDQSIKVRISLKSALGEKA  632
Paguma         VVGEKKMNVTPLLKYFEPLFTWLKEQNRNSFVGWNTDWRPYADQSIKVRISLKSALGDEA  632
Felis          VVGEKNMNVTPLLKYFEPLFTWLKEQNRNSFVGWNTDWRPYADQSIKVRISLKSALGDKA  624
Panthera       VVGEKNMNVTPLLKYFEPLFTWLKEQNRNSFVGWNTDWRPYADQSIKVRISLKSALGDKA  624
Manis          VVGTKNMDVRPLLNYFEPLLTWLKEQNRNSFVGWNTDWSPYADQSIKVRISLKSALGEKA  632
Equus          IVGVKNMDVRPLLNYFEPLFTWLKDQNKNSFVGWSTNWSPYADQSIKVRISLKSALGEKS  632
               :.. .*:. *: :**:**  **:.  * :  ***..:* **: ::*********:***  .:

Ophiophagus    YNWDESEMFLFKSTIAYAMQKYFLEVKNKTVPFH---------------------------  664
Mesocricetus   YEWDDNEMYLFRASVAYAMRVYFAKNKTQTVPFGVEDIRVSDLKPRVSFNFFVTSPQNVS  692
Mus            YEWTNNEMFLFRSSVAYAMRKYFSIIKNQTVPFLEEDVRVSDLKPRVSFYFFVTSPQNVS  692
Rhinolophus    YEWNDNEMYLFRSSVAYAMRKYFSNLKPRISFNFHVTSPGNLS  692
Bos            YEWNDNEMYLFQSSVAYAMRKYFSEARNETVLFGEDNVWVSDKKPRISFKFFVTSPNNVS  691
Homo           YEWNDNEMYLFRSSVAYAMRQYFLKVKNQMILFGEEDVRVANLKPRISFNFFVTAPKNVS  692
Macaca         YEWNDNEMYLFRSSVAYAMRTYFLEIKHQTILFGEEDVRVADLKPRISFNFYVTAPKNVS  692
Mustela        YEWNDNEMYFFQSSIAYAMREYFSKVKNQTIPFVGKDVRVSDLKPRISFNFIVTSPENMS  692
Canis          YEWNDNEMYLFRSSIAYAMRQYFSKVKNQTIPFVEDNVRVSDLKPRISFNFFVTSPGNVS  691
Paguma         YEWNDNEMYLFRSSIAYAMREYFSKVKNQTIPFVEDNVVSDLKPRISFNFFVTFSNNVS  692
Felis          YEWNDNEMYLFRSSVAYAMREYFSKVKNQTIPFVEDNVWVSNLKPRISFNFFVTASKNVS  692
Panthera       YEWNDNEMYLFRSSVAYAMREYFSKVKNQTIPFVEDNVWVSNLKPRISFNFFVTASKNVS  684
Manis          YEWNDSEMYLFRSSVAYAMREYFSKVKKQTIPFEDECVRVSDLKPRVSFIFFVTLPKNVS  692
Equus          YEWNDNEMYLFQSSVAYAMRVYFLKAKNQTILFGEEDVWVSDLKPRISFNFFVTSPKNAS  692
               *:*  :.**::**:::****:  **        : : : *

Ophiophagus    --------------LSRDRINEAFKLTDQTLEFIGLLPTLAPPYESPITVWLVAFGVVIGL  711
Mesocricetus   DIIPRNEVEAVRLSRGRINDVFGLDDNSLEFLGINPTLSPPYQPPVTIWLIFFGVVMGI  752
Mus            DVIPRSEVEDAIRSRGRINDVFGLNDNSLEFLGIHPTLEPPYQPPVTIWLIIFGVVMAL  752
Rhinolophus    DIIPRPEVEGAIRMSRSRINDAFRLDDNSLEFLGIQPTLGPPYQPPVTIWLIVFGVVMAV  752
Bos            DIIPRTEVENAIRLSRDRFNDAFRLNDNSLEFLGIQPTLGPPYQPPVTIWLIIFGVVMGV  751
Homo           DIIPRTEVEKAIRMSRSRINDAFRLNDNSLEFLGIQPTLGPPNQPPVSIWLIVFGVVMGV  752
Macaca         DIIPRTEVEEAIRISRSRINDAFRLEDNSLEFLAPPYQSPVTIWLIVFGVVMGV  752
Mustela        DIIPRADVEEAIRKSRGRINDAFRLDDNSLEFLGIQPTLEPPYQPPVTIWLIVFGVVMGV  752
Canis          DIIPRTEVEEAIRMYRSRINDVFRLDDNSLEFLGIQPTLGPPYEPPVTIWLIVFGVVMGV  751
Paguma         DVIPRSEVEDAIRMSRSRINDAFRLDDNSLEFLGIEPTLSPPYRPPVTIWLIVFGVVMGA  752
Felis          DVIPRSEVEEAIRMSRSRINDAFRLDDNSLEFLGIQPTLSPPYQPPVTIWLIVFGVVMGV  752
Panthera       DVIPRREVEEAIRMSRSRINDAFRLDDNSLEFLGIQPTLSPPYQPPVTIWLIVFGVVMGV  744
Manis          AVIPRAEVEEAIRISRSRINDAFRLDDNSLEFLGIQPTLQPPYQPPVTIWLIVFGVVMGV  752
Equus          DIIPRTDVEEAIRMSRSRINDAFRLDDNTLEFLGIQPTLGPPYQPPVTVWLIAFGVVMGL  752
               .*.*:*:. * *.*:::***:* **     **.*** *  .*:: **. **:**::

Ophiophagus    IVIGIITLEKAGSKN-----------------------------------          726
Mesocricetus   VVVGIIILIFTGIKGRKKKKNETKREENPYDSVDIGKGESNAGFLSNDDAQTSF        805
Mus            VVVGIIIILIVTGIRNRKKKNETKREENPYDSMDIGKGESNAGFQNSDDAQTSF        805
Rhinolophus    VVVGIVVLIITGIRDRRKTDQARSEENPYSSVDLSKGENNPGFQNSDDVQTSF         805
Bos            VVIGIVIIFTGIRNRKKQASSEENPYASVDLNKGENNSGFVNIDDVQTSL            804
Homo           IVVGIVILIFTGIRDRKKKNKARSGENPYASIDISKGENNPGFQNTDDVQTSF         805
Macaca         IVAGIVVILIFTGIRDKKNQARSEENPYASIDINKGENNPGFQNTDDVQTSF          805
Mustela        VVVGIFLLIFSGIRNRKKNDQARGEENPYASVDLSKGENNPGFQNVDDAQTSF         805
Canis          VVVGIVLLIFSGIRNRRKNDQARGEENPYASVDLSKGENNPGFQNVDDAQTSF         804
Paguma         IVVGIVLLIVSGIRNRRKNDQAGSEENPYASVDLNKGENNPGFQHADDVQTSF         805
Felis          VVVGIVLLIVSGIRNRRKNNQARSEENPYASVDLSKGENNPGFQHADDVQTSF         805
Panthera       VVVGIVLLIVSGIRNRRKNNQARSEENPYASVDLSKGENNPGFQHADDVQTSF         797
Manis          VVVGIVLLIFKDQARSEENPYASVDLSKGENNPGFQNVDDVQTSF                 805
Equus          VVVGIVVLIATGIRGRRKNQARSEENPYASVDLSKGENNPGFQNGDDVQTSF          805
               :* **. * :*  :.
```

**Supplementary Figure 2.** Sequence alignment of ACE2 amino acid sequence from selected species. The SAR-Cov-2 spike protein binding region is highlighted in red.

**BAT ACE2**

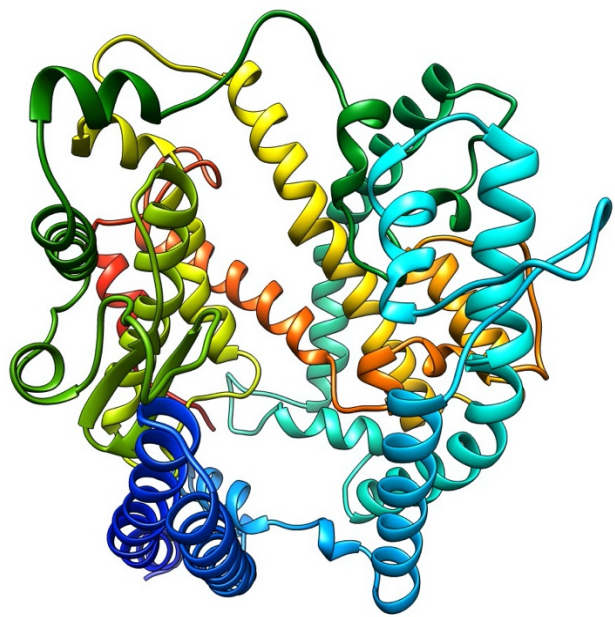
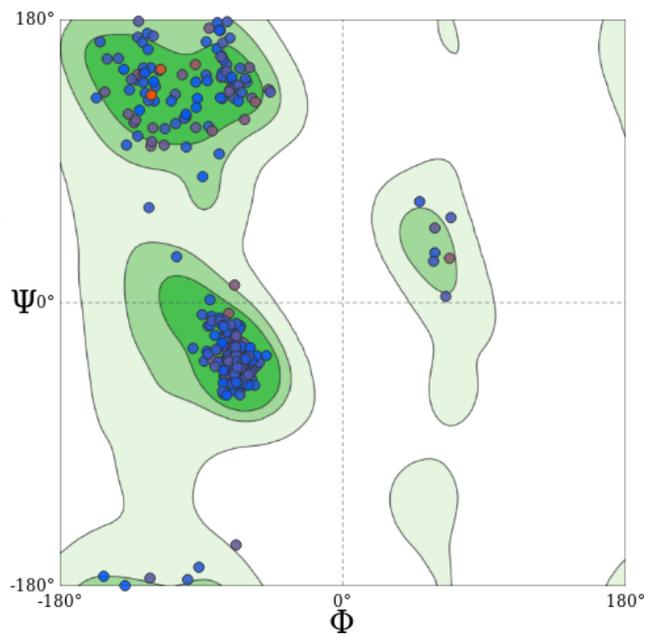
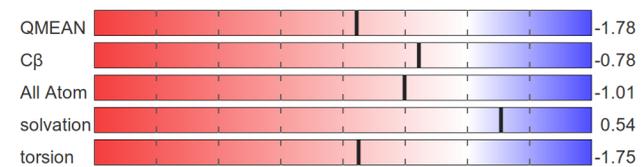
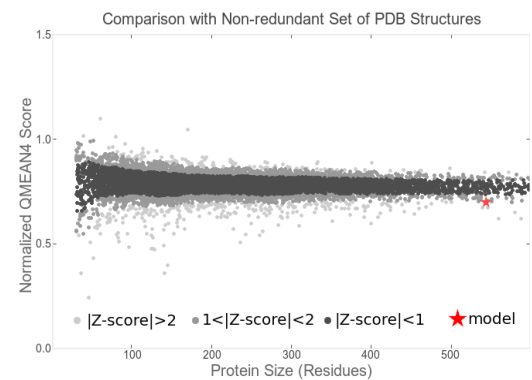

**CAT ACE2**

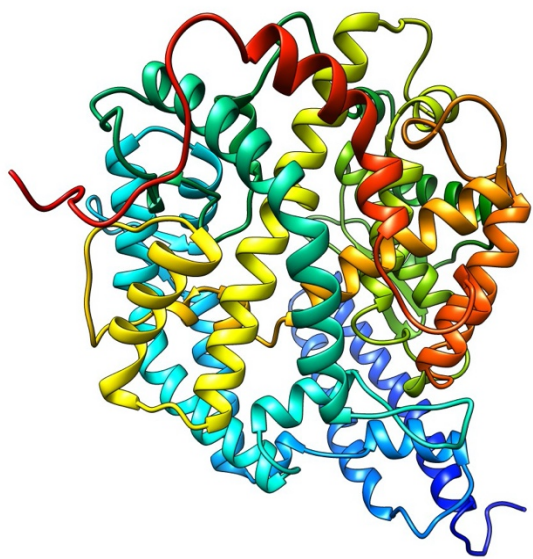
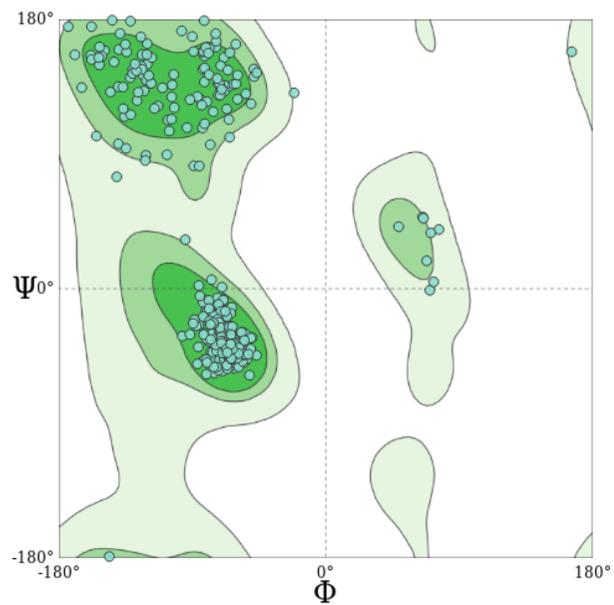
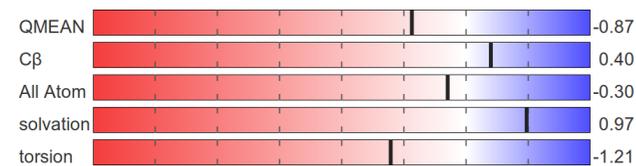
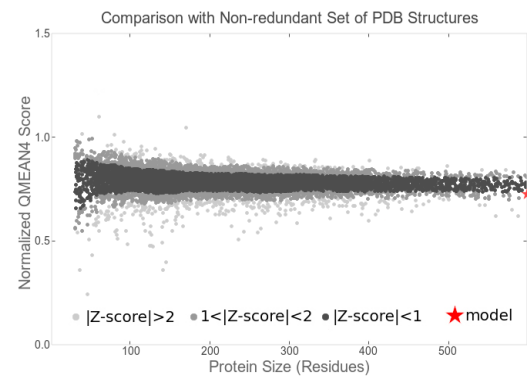

**CIVET ACE2**

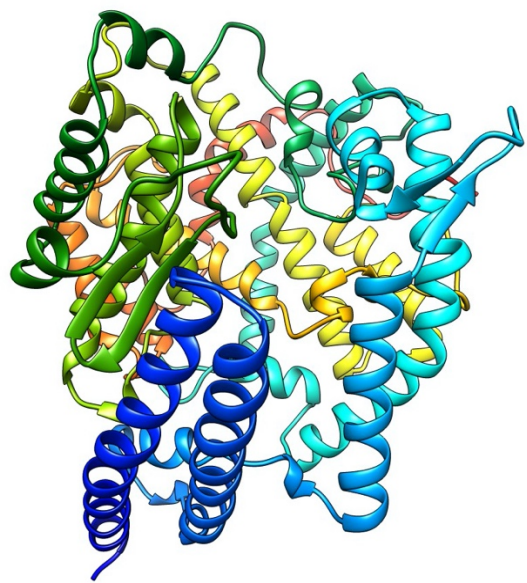
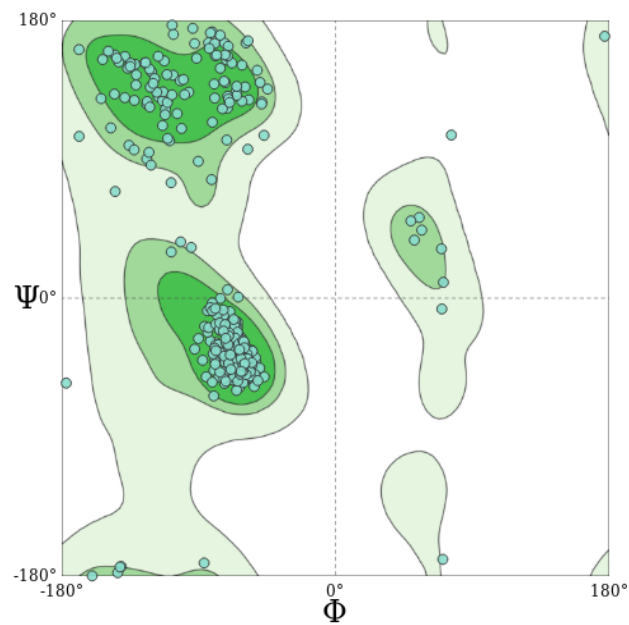
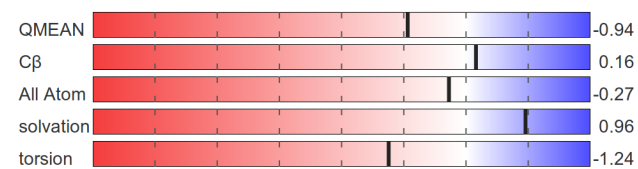
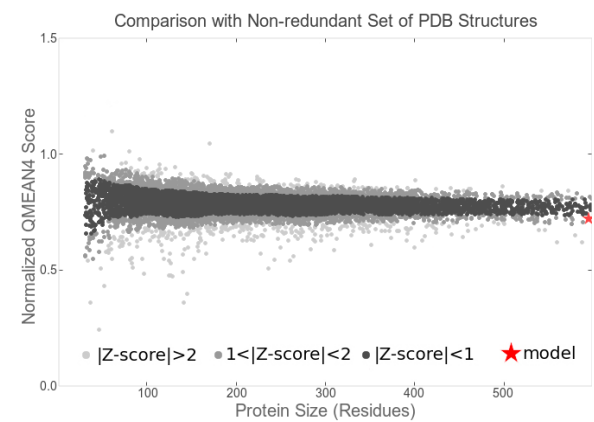

**DOG ACE2**

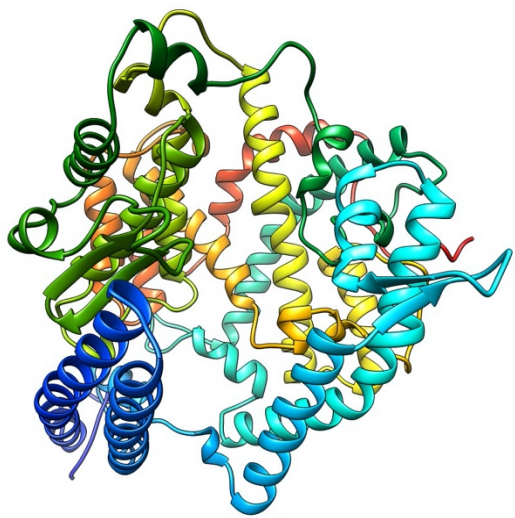
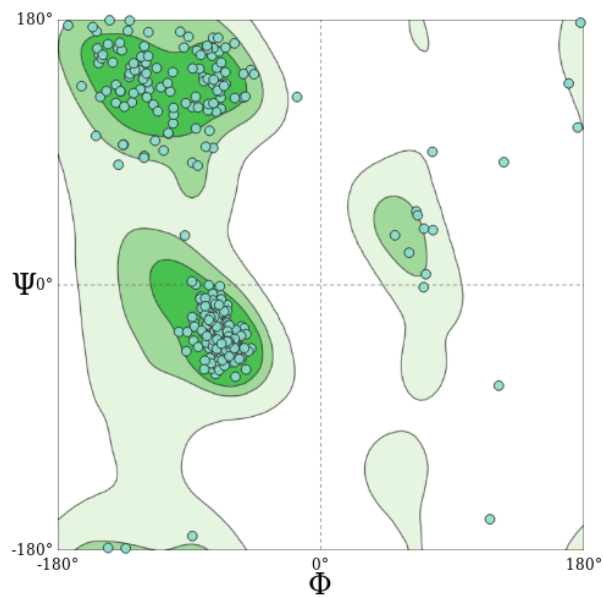
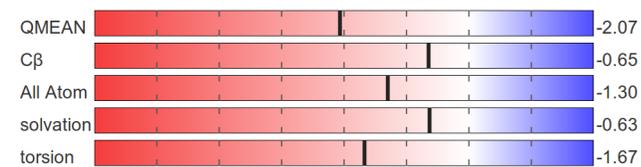
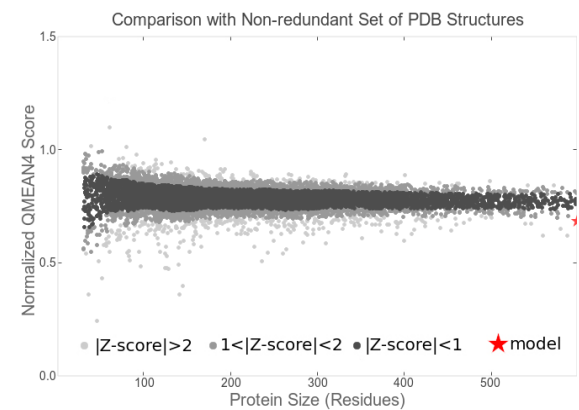

**FERRET ACE2**

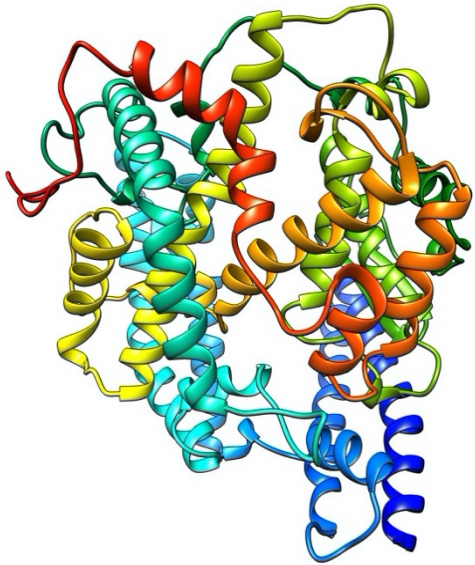
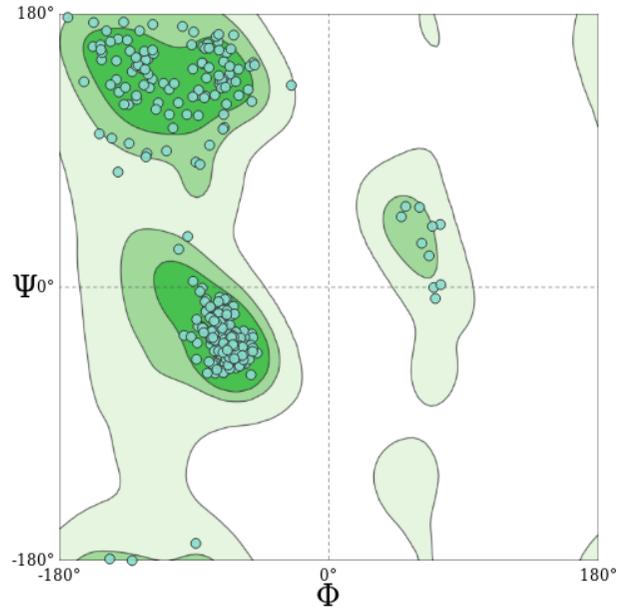
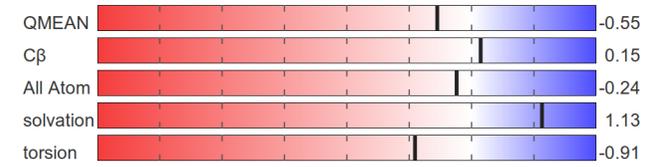
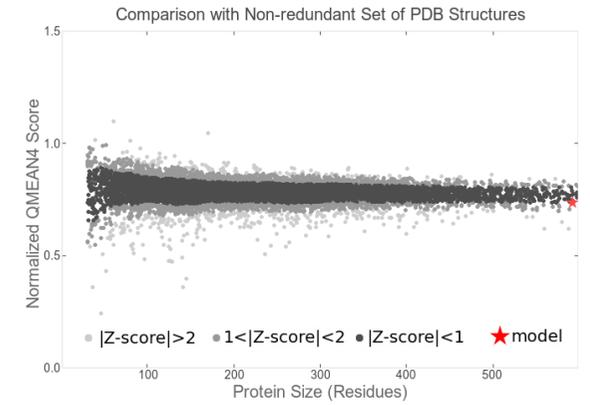

**HAMSTER ACE2**

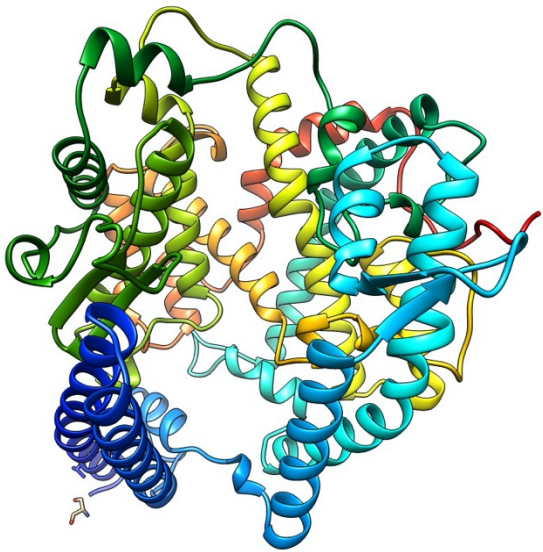
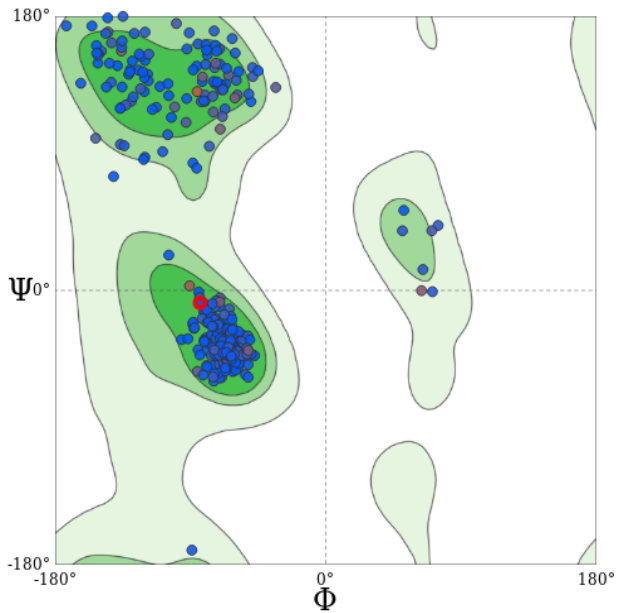
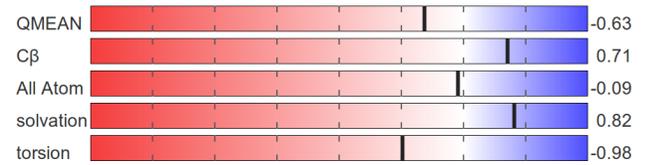
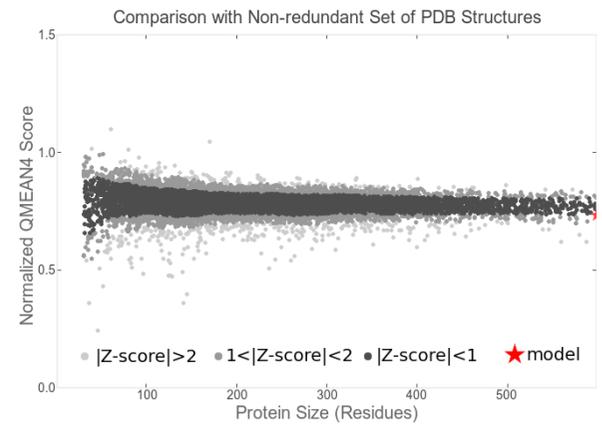

**MONKEY ACE2**

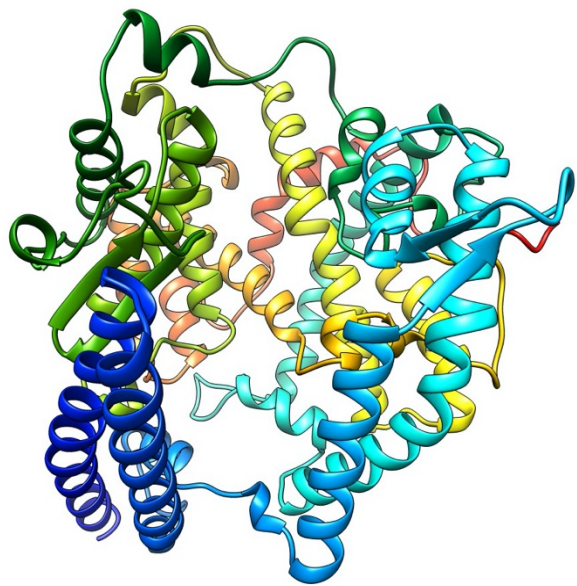
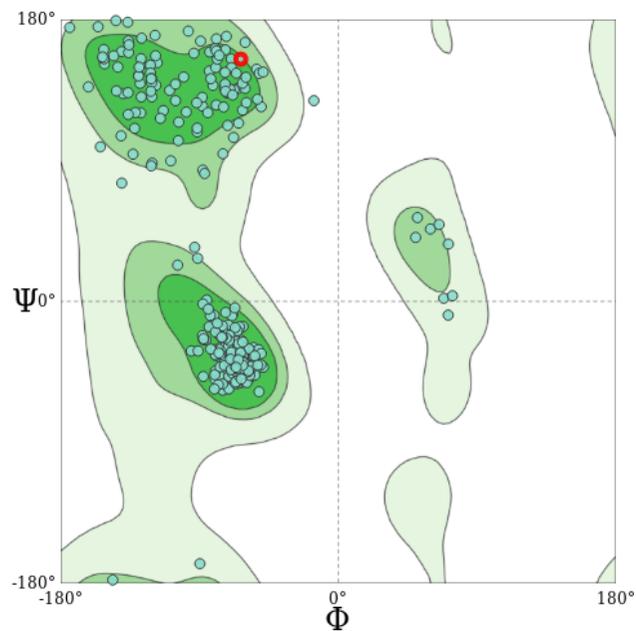
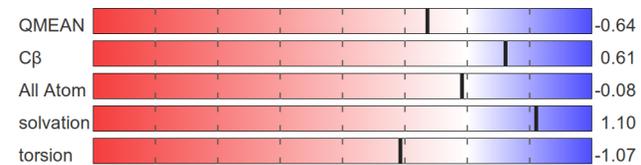
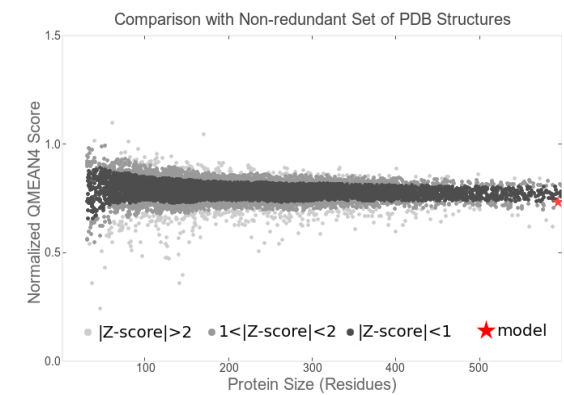

**MOUSE ACE2**

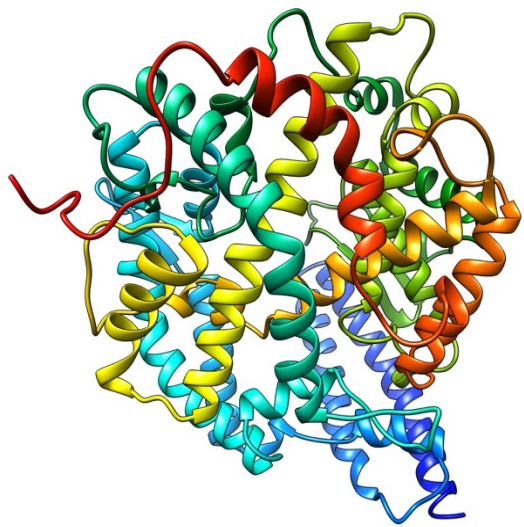
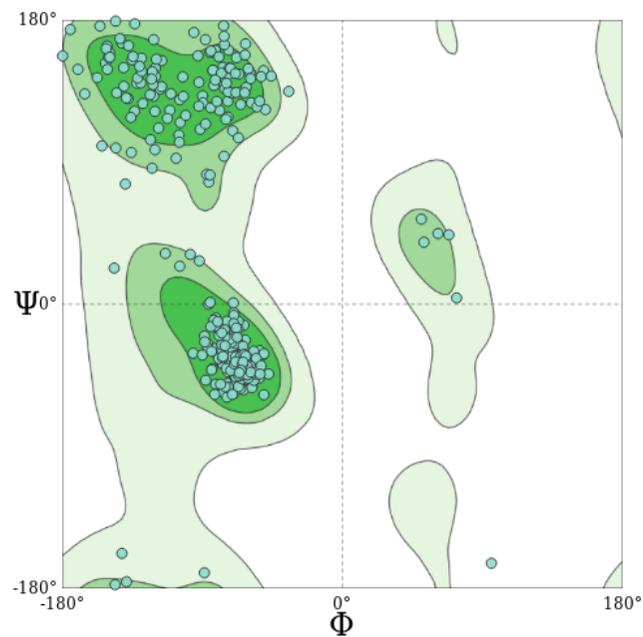
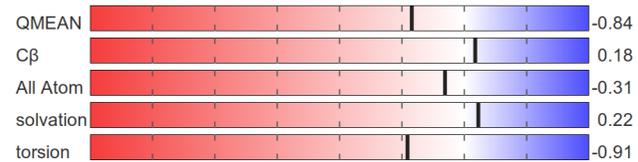
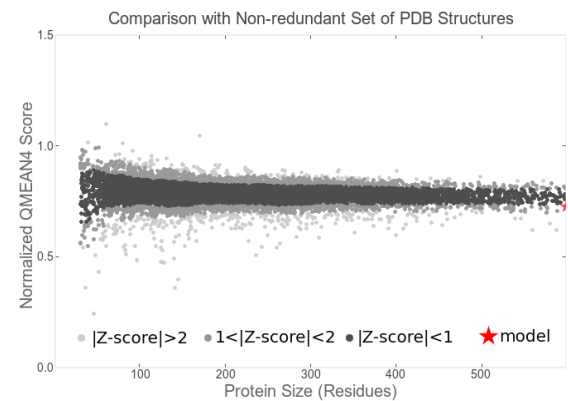

**PANGOLIN ACE2**

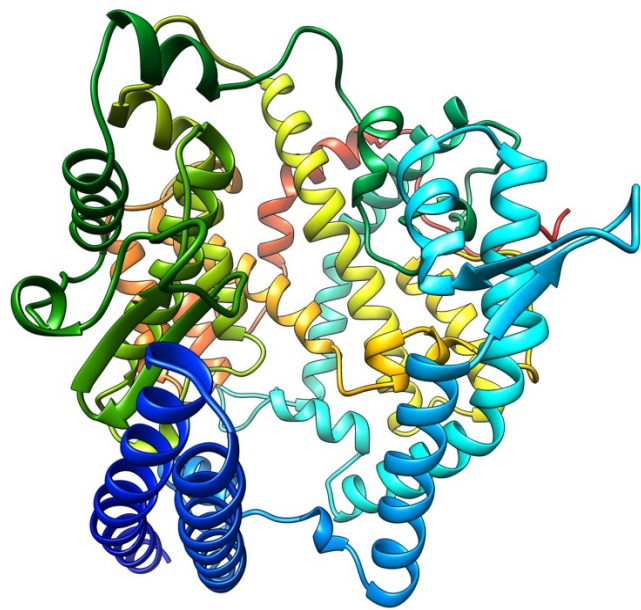
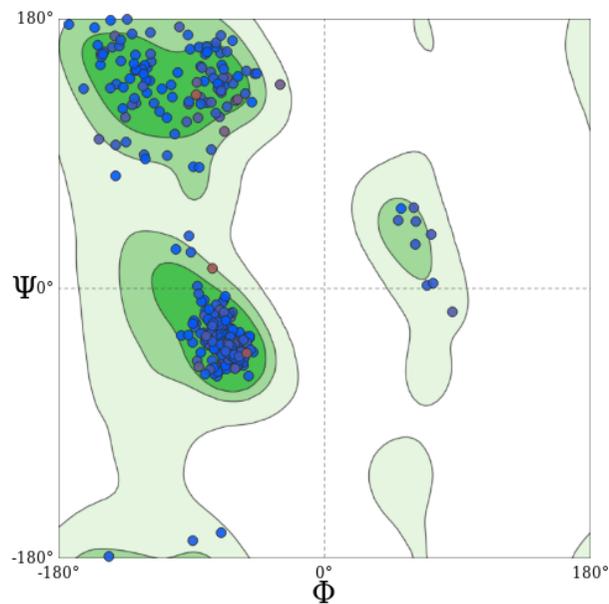
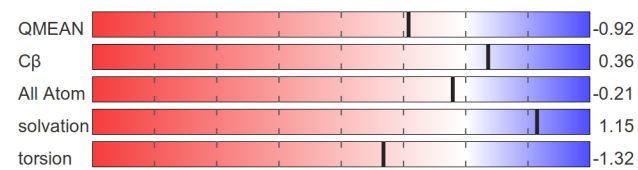
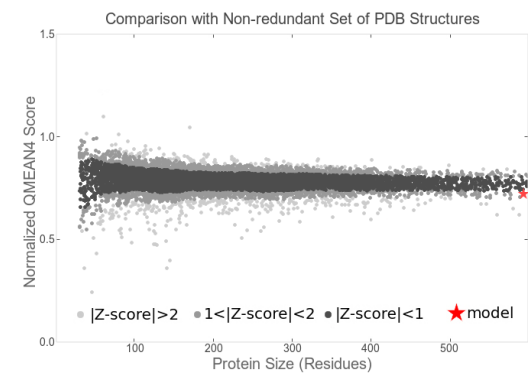

**SNAKE ACE2**

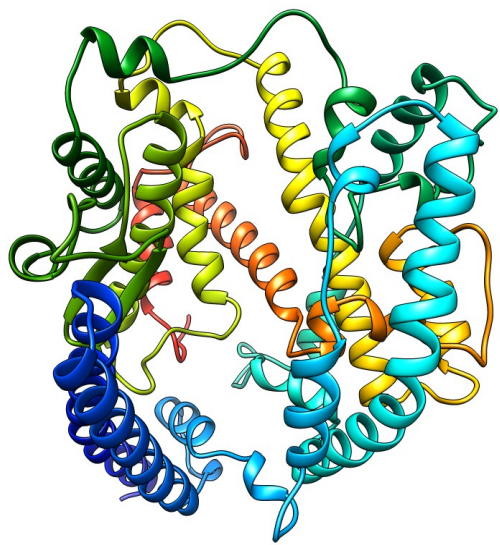
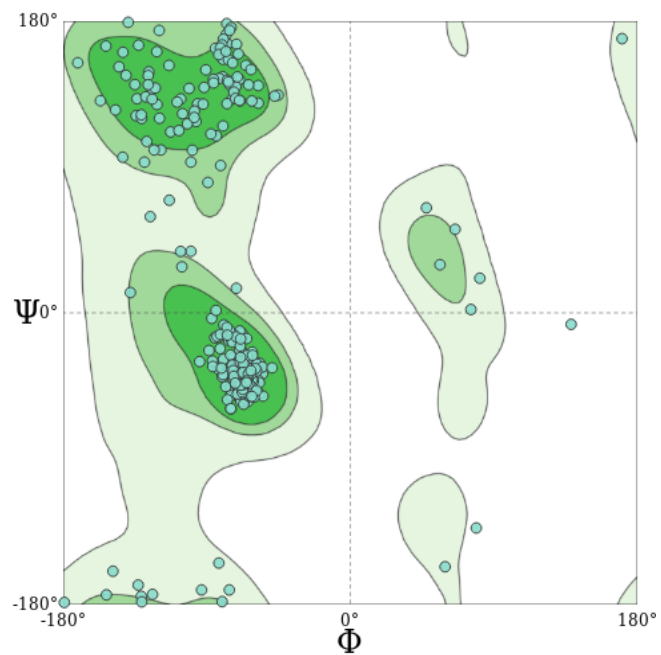
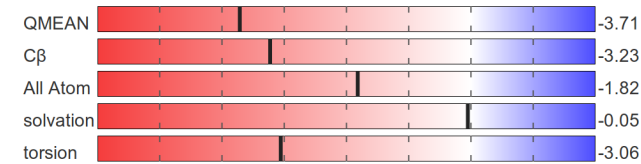
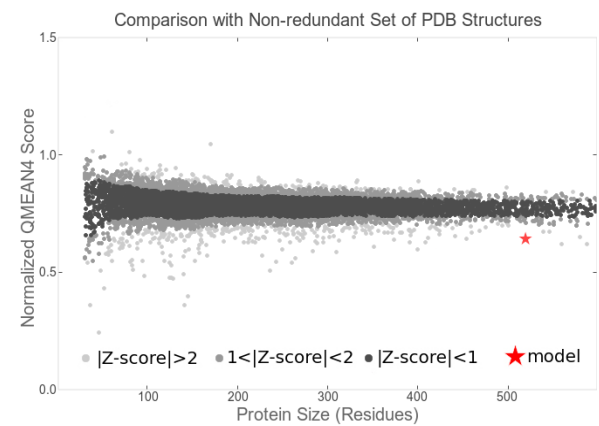

**HORSE ACE2**

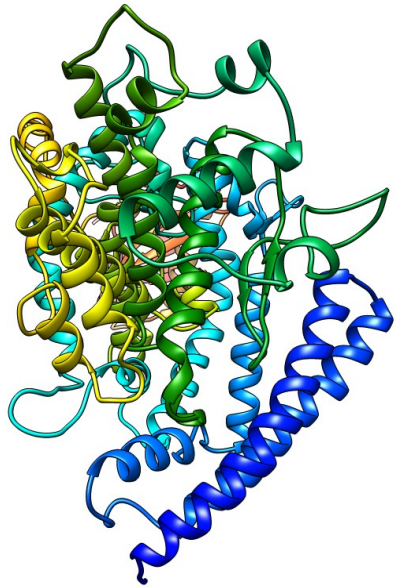
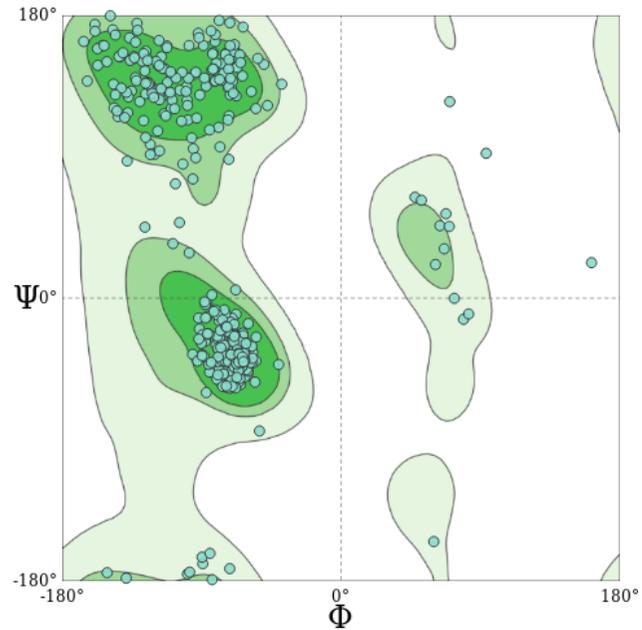
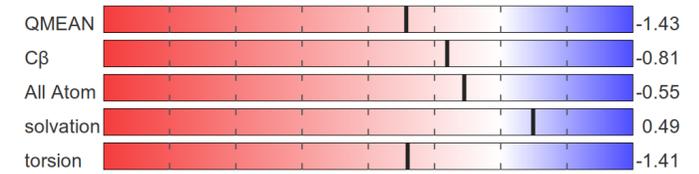
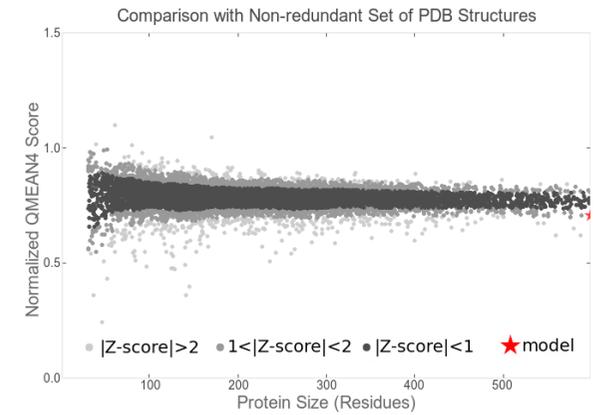

**CATTLE ACE2**

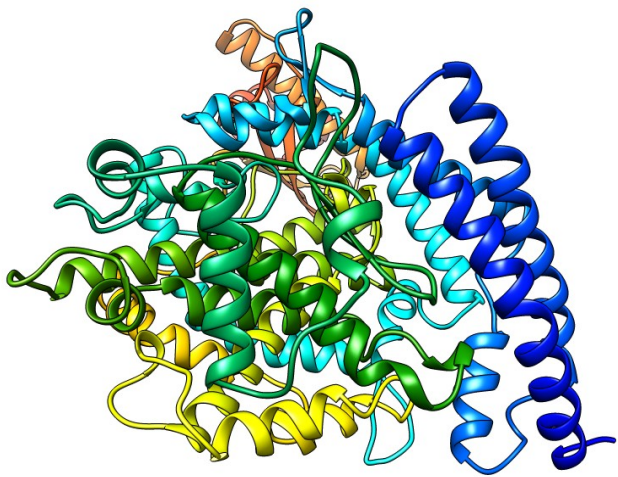
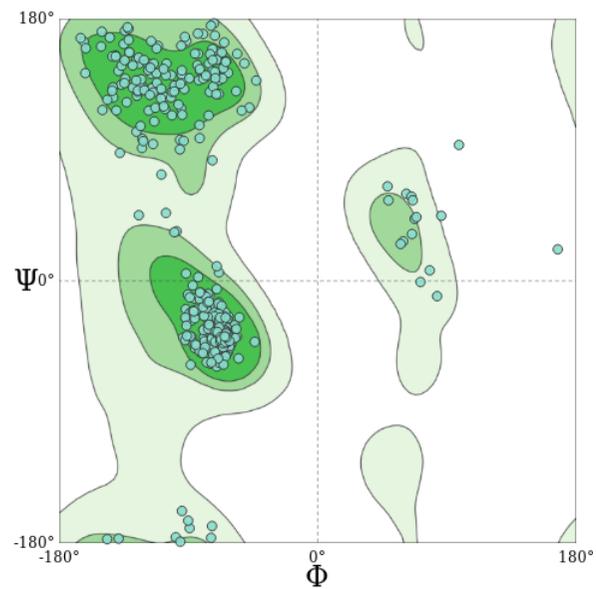
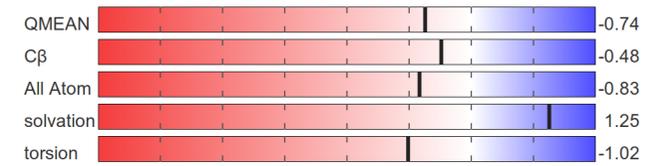
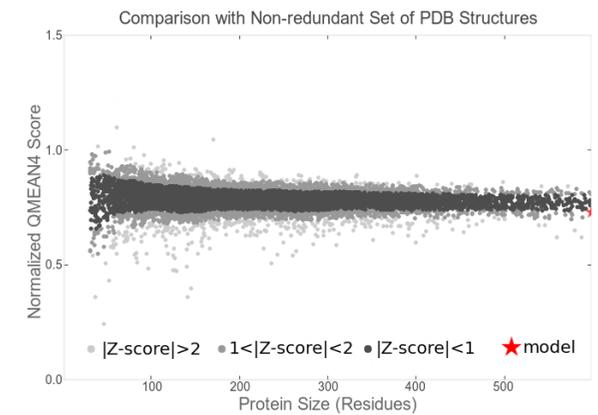

# TIGER ACE2

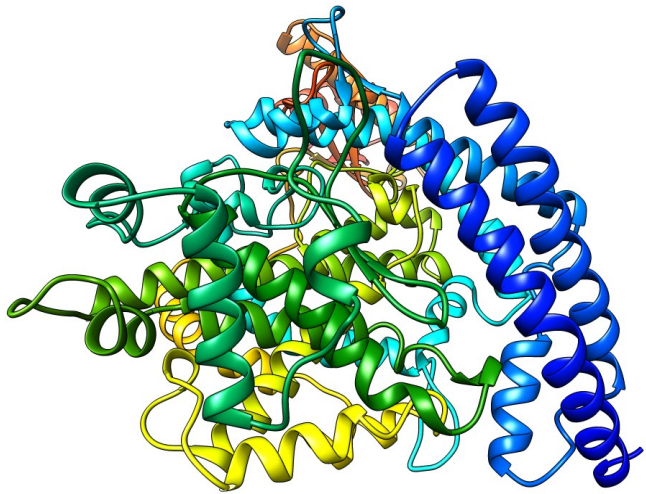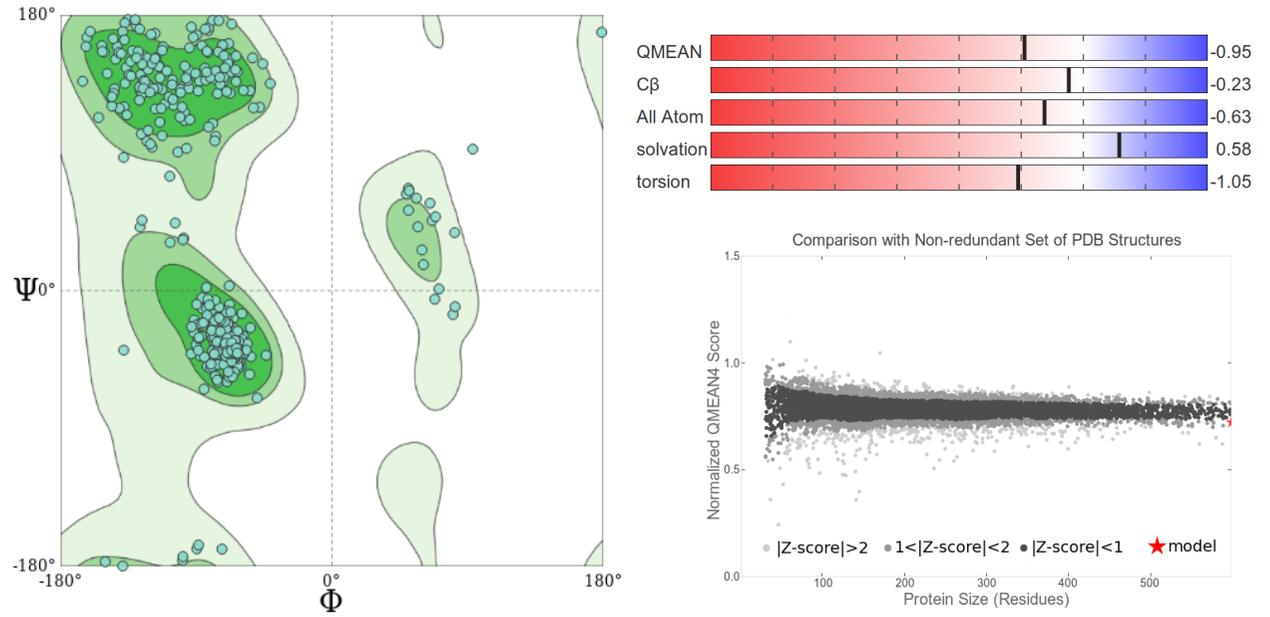

**Supplementary Figure 3**. Modelled ACE2 structures for selected species, with Ramachandran plots and quality met

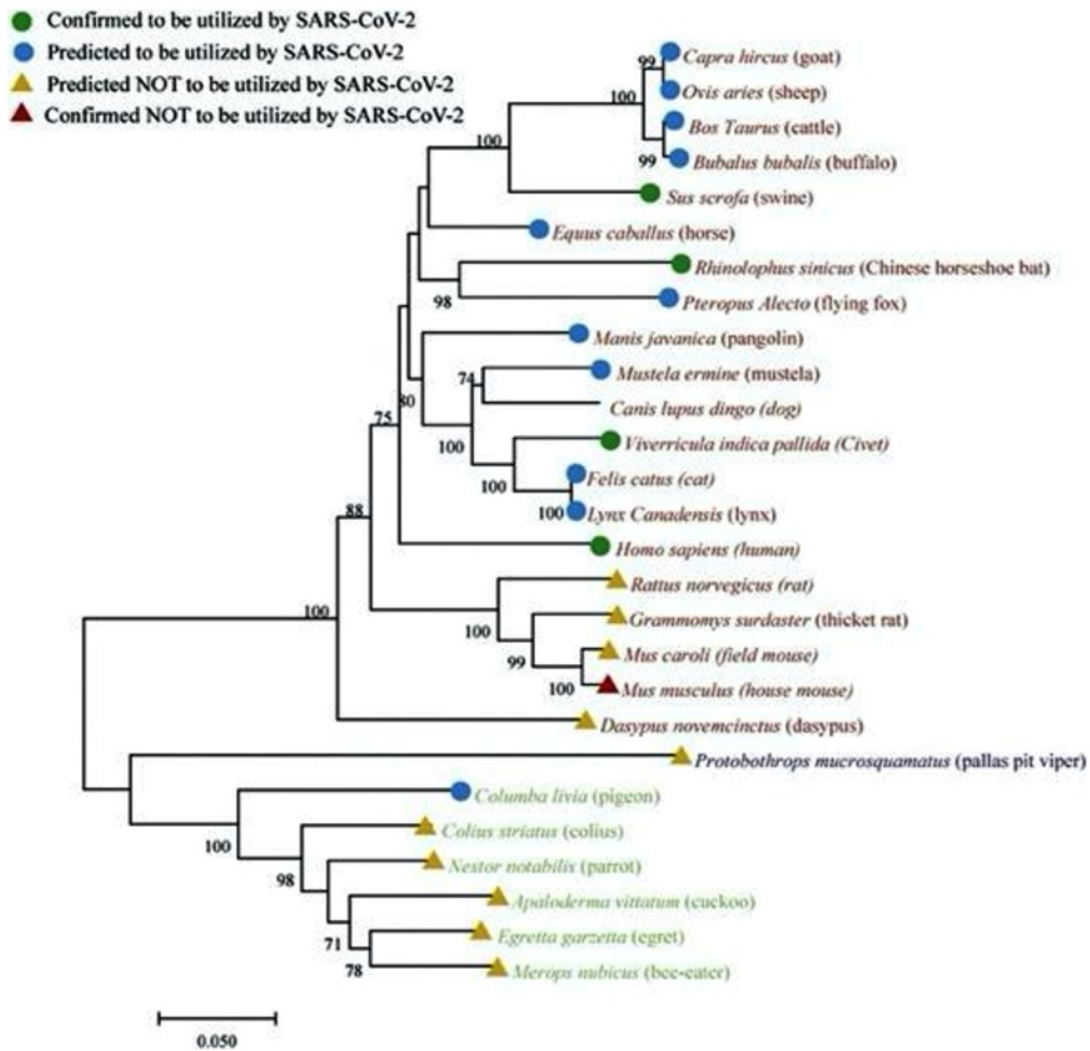

**Supplementary Figure 4**. Predicted and confirmed utilization of ACE2 receptors by the SARS-Cov-2 spike protein based on sequence homology from Qui et al.[1]

**References**

1    Qiu, Y. *et al.* Predicting the angiotensin converting enzyme 2 (ACE2) utilizing capability as the receptor of SARS-CoV-2. *Microbes Infect.* **22**, 221-225, doi:10.1016/j.micinf.2020.03.003 (2020).